\documentclass[lettersize,journal]{IEEEtran}
\usepackage{amsmath,amsfonts}
\usepackage{algorithmic}
\usepackage{algorithm}
\usepackage{array}
\usepackage{textcomp}
\usepackage{stfloats}
\usepackage{url}
\usepackage{verbatim}
\usepackage{graphicx}
\usepackage{caption}
\usepackage{romannum}
\usepackage{cite}
\usepackage{amsmath}
\usepackage{makecell}
\usepackage{multirow}
\usepackage{subfigure}
\usepackage{xcolor}
\usepackage{hyperref}
\usepackage{subcaption}

\begin{document}

\title{Bayesian Bias Learning-Enabled Probabilistic Voltage Estimation with Unknown Uncertainty Modeling of DERs in Distribution Grids}

\author{
        Yuanshuo Zhang,~\IEEEmembership{Graduate Student Member,~IEEE},      
        Ying Zhang,~\IEEEmembership{Senior Member,~IEEE},   
        Fei Ding,~\IEEEmembership{Senior Member,~IEEE},
        Jiahao Chen,~\IEEEmembership{Graduate Student Member,~IEEE} 

\thanks{Y. Zhang, Y. Zhang, and J. Chen are with the School of Electrical and Computer Engineering, Oklahoma State University, Stillwater, OK 74074, USA (e-mail: yuanshuo.zhang@okstate.edu; y.zhang@okstate.edu; jiahao.chen@okstate.edu).

F. Ding is with the Power Systems Engineering Center, National Laboratory of the Rockies, Golden, CO 80401, USA (e-mail: Fei.Ding@nlr.gov).} }
% This research was supported by the National Science Foundation through award numbers 2346213 and 2418359.

\maketitle
\begin{abstract}
One key challenge in distribution system state estimation is unknown input statistics from the growing integration of distributed energy resources (DERs), which may largely deviate from Gaussian uncertainty. Existing model-based estimators often rely on the assumption that all inputs follow known Gaussian distributions, while data-driven methods suffer from overfitting and failure in out-of-distribution scenarios.
This paper proposes a Bayesian bias learning (BBL)-enabled probabilistic voltage estimation method in distribution systems with unknown DER generation statistics. The proposed learning-enabled method consists of two stages: probabilistic uncertainty inference and computationally efficient voltage estimation.
First, a physics-constrained probabilistic density inference network is developed to learn Gaussian mixture representations of DER generation from historical data. Then, the resulting probabilistic representations, treated as pseudo-measurements, are integrated into a BBL-enabled voltage estimator via a decomposition-and-parallelization mechanism. Furthermore, utilizing a linear analytical solution, the proposed BBL-enabled voltage estimator efficiently captures linearization-induced biases and attains accurate probabilistic estimation. 
By integrating learning-enabled probabilistic uncertainty modeling into the parallel estimation structure, the proposed method achieves uncertainty-aware scalable voltage estimation.
Case studies on two unbalanced IEEE distribution networks show that the proposed estimator achieves significantly higher accuracy and computational efficiency under unknown uncertainty, outperforming recent data-driven and model-based methods.
\end{abstract}


\section{Introduction}
\IEEEPARstart{D}{istribution} system state estimation (DSSE), which is the heart of advanced distribution management systems, attains real-time system-wide status from measurement data
and available information. 
Because of the scarcity of real-time measurements such as phasor measurement units (PMUs), state estimation in low-voltage distribution systems relies heavily on pseudo-measurements \cite{pse7312455} to maintain observability. The pseudo-measurements are derived from historical or predicted load and distributed energy resource (DER) generation, with a large intricate uncertainty associated. However, the probabilistic characteristics of DER generation or loads significantly deviate from Gaussian behavior \cite{singh2010disGMM,zhangInterval,hoeltgebaum2018generating}. These unknown non-Gaussian uncertainties render the Gaussian noise assumptions underlying traditional weighted least squares (WLS) estimators invalid. To maximize the DSSE performance, it is important to design flexible state estimators able to deal with arbitrary kinds of uncertainty distributions.

% \textcolor{blue}{Adaptive filtering has also been investigated for voltage estimation under measurement disturbances. For direct current-voltage estimation, a robust Laplace continuous mixed-norm adaptive filter is introduced to improve estimation stability under impulsive noise \cite{zayyani2024robust}. Nevertheless, these adaptive filtering methods.}

Recent studies have incorporated exact modeling of various uncertainties into DSSE formulations as precise pseudo-measurements.
Gaussian mixture models (GMMs) are used to formulate arbitrary distributions as weighted sums of Gaussian components \cite{5298967GMM}. 
 In \cite{NonGaussianvanin2023exact}, the exact probability function of various types of load measurement errors, e.g., normal, log-normal, and beta distributions, is integrated in the estimation function.  
References \cite{NonGaussianvanin2023exact} and \cite{NonGaussian8008784} develop maximum likelihood estimation (MLE) and Bayesian inference to accommodate non-Gaussian input uncertainty. In \cite{zayyani2024robust}, a Laplace continuous mixed norm adaptive filter method is developed in DC voltage estimation by incorporating an exponential weighting function.
The performance of most model-based algorithms relies heavily on specific statistical distributions and DER uncertainty modeling accuracy; any approximation used to represent this uncertainty in the DSSE model translates into errors that propagate through the network. 
To enhance adaptability, machine learning (ML)-based DSSE approaches have been developed when an accurate grid or uncertainty models are unavailable. These methods can be mainly divided into two categories: direct state inference \cite{zamzam2019data,liu2025noise,JZhTPWRS10184461,azimian2022state,tran2021,zamzam2020,yuan2022joint,caodi2024tpwps}, and probabilistic uncertainty modeling-aided estimation \cite{manitsas2012distribution,huang2019probabilistic,dobbe2019linear, pse8616788,mestav2019bayesian}. A more granular taxonomy is given in Table \ref{tab:lit_comp}.

\begin{table}[!t]
\centering
\caption{\textsc{Taxonomy of ML-Based DSSE Methods}}
\label{tab:lit_comp}
\small
\begin{tabular}{|c|c|c|}
\hline
\textbf{Category} 
& \textbf{Reference} 
& \makecell{\textbf{Handle Unkonwn} \\ \textbf{Non-Gaussian}\\\textbf{Uncertainty?}} \\ 
\hline
\multirow{3}{*}{\makecell{Direct State Inference}}
& \cite{zamzam2019data,liu2025noise,JZhTPWRS10184461}
& $\times$ \\ \cline{2-3}& \cite{azimian2022state}& $\checkmark$ \\ 
\cline{2-3}
& \cite{tran2021,zamzam2020,yuan2022joint,caodi2024tpwps}& $\times$, physics-informed\\ 
\hline
\multirow{3}{*}{\makecell{Probabilistic Uncertainty\\Modeling-aided Estimation}}
& \cite{manitsas2012distribution,huang2019probabilistic} & $\checkmark$ \\ \cline{2-3}& \cite{dobbe2019linear,pse8616788,mestav2019bayesian} 
& $\times$ \\ \cline{2-3}& Proposed& $\checkmark$, physics-informed \\ 
\hline
\end{tabular}
\end{table}
In the first category, ML-based implementations directly map the noisy measurements to voltage states by
minimizing a predetermined loss function over the training data. For example, \cite{zamzam2019data} trains a shallow neural network to obtain an initial estimate near the true voltage and initializes the iterative Gauss–Newton solver.
% Reference \cite{he2020power} employs a generative adversarial network (GAN) by learning latent data distributions, and \cite{liu2025noise} extends it to handle extreme and missing data scenarios.
A multi-fidelity deep Gaussian process is developed in \cite{JZhTPWRS10184461} to infer voltage states from heterogeneous measurement sources, including PMUs and pseudo-measurements. The robustness of deep neural network (DNN)-based voltage estimation is evaluated in \cite{azimian2022state} by injecting non-Gaussian noise into measurements.
 However, the above methods fall under supervised learning, most of which rely heavily on a large amount of high-quality training data and suffer from overfitting \cite{raghuvamsi2023review}.
 To mitigate these issues, various physics-informed ML-based DSSE methods have emerged recently. For example, \cite{tran2021} aligns the DNN structure with the actual topology by pruning redundant connections that violate physical connectivity. \cite{zamzam2020} and \cite{caodi2024tpwps} embed the information of network topology or PMU locations into the neural network architecture. 
 A hierarchical reinforcement learning-aided framework is developed in \cite{yuan2022joint} to enhance online computational efficiency, which trains actor-critic agent models for voltage estimation in the secondary distribution grid while retaining the WLS-based estimator for the primary network.
 However, \cite{tran2021,zamzam2020,yuan2022joint,caodi2024tpwps} do not consider non-Gaussian measurement uncertainties. 

Motivated by the need for dedicated uncertainty quantification, uncertainty modeling-aided DSSE methods \cite{manitsas2012distribution, huang2019probabilistic, dobbe2019linear,pse8616788,mestav2019bayesian} are developed, where the probabilistic distribution of uncertain sources is inferred, explicitly modeled, and integrated into a state estimator. Following this idea, pseudo-measurements are generated by DNNs in \cite{manitsas2012distribution} and deep belief networks (DBNs) in \cite{huang2019probabilistic}, with the uncertainty parameters of pseudo-measurements modeled as GMMs. Although \cite{manitsas2012distribution, huang2019probabilistic} improve DSSE performance under non-Gaussian uncertainty, their state estimation stages retain iterative runs and incur prohibitive computational time as the number of complex uncertainty sources grows. Such computational burdens limit the scalability for real-time system-wide monitoring with massive DER installation. A linear least-squares estimator for unbalanced distribution systems is formulated in \cite{dobbe2019linear}, where a Gaussian process is first employed to construct prior voltage predictions for unmeasured nodes. Reference \cite{pse8616788} develops parallel relevance vector machines and game-theoretic weighting to generate pseudo-measurements and uncertainty estimates for seasonal load patterns. Nevertheless, stochastic variations in power generation due to DER penetration can considerably degrade the DSSE performance \cite{dobbe2019linear}. To address this challenge, \cite{mestav2019bayesian} learns the distribution of stochastic power injections as prior information and embeds it into a DNN-based voltage estimator to enforce the minimum mean-square error. However, the uncertainty modeling in \cite{dobbe2019linear, pse8616788,mestav2019bayesian} is restricted to Gaussian assumptions and point-wise estimation and cannot model uncertainty distributions and non-Gaussian characteristics. In summary, most data-driven methods face several limitations, including heavy reliance on massive high-quality training datasets and a lack of robustness against out-of-distribution scenarios during online execution.

To address the aforementioned challenges under unknown non-Gaussian DER uncertainty, this paper proposes a Bayesian bias learning (BBL)-enabled probabilistic voltage estimation algorithm in unbalanced distribution systems. The proposed learning-enabled method, which falls under physics-informed/guided ML, consists of two stages: probabilistic uncertainty inference and computationally efficient voltage estimation.
First, a physics-constrained probabilistic density inference network (PDIN) is proposed to jointly generate pseudo-measurements and their probabilistic distributions in the generalized GMM form, which is expressed as a combination of multiple Gaussian components. Then, a BBL-enabled decomposition-and-parallelization mechanism is developed to form multiple parallel voltage estimation subproblems.
For each subproblem, based on a linear analytical solution, the proposed BBL-enabled voltage estimator is offline trained to capture linearization-induced biases for accurate probabilistic estimation. 
By integrating probabilistic uncertainty modeling with the learning-enabled parallel estimation structure, the proposed method achieves accurate, uncertainty-aware, and scalable voltage estimation under unknown input statistics. 
The major contributions are listed below:
\vspace{-1pt}
\begin{itemize}
\item Compared with data-driven pseudo-measurement generation methods using Gaussian assumptions, the proposed PDIN-based method jointly produces probabilistic pseudo-measurements in the form of GMMs to model unknown uncertainty by a dual-branch structure. Moreover, its novel development of physics-constrained activation and multi-task loss functions enforces statistical significance on the GMM parameters and addresses the density leakage issue inherent in conventional single-task density learning networks.
\item Within the decomposition-and-parallelization mechanism, the proposed BBL-enabled voltage estimation enables scalable probabilistic uncertainty quantification. 
The proposed stage-wise method effectively mitigates the limitations of direct Bayesian learning, which favors Gaussian priors and struggles to handle arbitrary uncertainties.
\item Unlike the purely deep learning method, the proposed BBL-enabled surrogate, which is physics-guided, preserves the mathematical relationships between measurements and system states, thereby enhancing interpretability and robustness in learning. This hybrid method achieves a synergy between conventional linear approximation models and direct data-driven DSSE methods.
\item The proposed method is compared with probabilistic learning-based uncertainty modeling methods, as well as model-based WLS and recent residual learning-driven DSSE methods. Case studies show that the proposed probabilistic approach achieves accurate, uncertainty-aware voltage estimation with superior computational efficiency. The sensitivity analysis shows that, 
% although the number of Gaussian-component combinations increases exponentially with the number of non-Gaussian uncertainty sources,
the learning-enabled parallel implementation effectively mitigates the resulting computational burden as the number of non-Gaussian uncertainty sources increases, enabling the proposed method to maintain high estimation accuracy with computational efficiency.
\end{itemize}

\section{Preliminary and Problem Formulation}
\subsection{Distribution System State Estimation}

In state estimation of an electric network, available measurements from various sensors across the grid are linked to the state variables by the following equation:
\begin{equation} \label{se}
\mathbf{z} =\mathbf{h}\left(\mathbf{x} \right) + \mathbf{e} \
\end{equation}
where $\mathbf{x} \in \mathbb{R}^n$ denotes the state vector, and $\mathbf{z} \in \mathbb{R}^m$ denotes the measurement vector, and $m>n$; $\mathbf{h}(\cdot)$ represents the known nonlinear measurement 
function w.r.t $\mathbf x$ and is determined by the distribution network model and measurement configuration \cite{abur2004power};
the measurement noise vector $\mathbf{e}$ is assumed to follow a known Gaussian distribution, i.e., $\mathbf{e} \sim \mathcal{N}(\mathbf{0}, \mathbf{R})$, 
where $\mathbf{R}$ denotes the measurement noise covariance matrix.

Define $\mathbf{x}$ as the bus voltage phasors in rectangular coordinates. 
For a multi-phase AC distribution system, the state vector is expressed as:
\begin{equation} \label{4_6} 
\mathbf{x}=[v^{\varphi}_{1r},v^{\varphi}_{1x},v^{\varphi}_{2r}, v^{\varphi}_{2x},v^{\varphi}_{3r}, v^{\varphi}_{3x},\cdots, v^{\varphi}_{N_br}, v^{\varphi}_{N_bx}] 
\end{equation} 
where ${v}^{\varphi}_{lr}$ and ${v}^{\varphi}_{lx}$ are the real and imaginary parts of the $\varphi$-phase voltage at bus $l$, $l=1,2,3,\cdots,N_b$. The superscripts $\varphi$ denote the phase indices.
Given that a limited number of high-accuracy PMUs are installed in distribution systems due to their high cost, hybrid state estimators incorporate supervisory control and data acquisition (SCADA) measurements and lower-accuracy pseudo-measurements alongside PMU data to improve observability. 

Therefore, the state variables are estimated by solving a WLS formulation, which minimizes the sum of weighted measurement residuals $J$, expressed as:
\begin{equation} 
\hat{\mathbf {x}} = {\rm{arg}}\;{\rm{min}}\;J = {\rm{arg}}\;{\rm{min}}{\left[ {\mathbf{z} - \mathbf{h}\left(\mathbf{x} \right)} \right]^T}\mathbf{W}\left[ {\mathbf{z} - \mathbf{h}\left(\mathbf{x} \right)} \right]
\label{objective}
\end{equation}
where $\mathbf{W}$ is the measurement weighting matrix that reflects the confidence levels of these heterogeneous measurements, and $\mathbf{W}=\mathbf{R}^{-1}$. 

Conventionally, the optimal state estimate is obtained iteratively using the Gauss-Newton method until the vector $\Delta \mathbf{x}$ becomes sufficiently small:
\begin{equation}\label{nonlinear2}
\frac{\partial \mathbf{J}}{\partial \mathbf{x}} = \mathbf{H}\left(\mathbf{x} \right)^T \mathbf{W} \left[ \mathbf{z} - \mathbf{h}\left(\mathbf{x} \right) \right] = 0
\end{equation}
\begin{equation}\label{nonlinear}
\mathbf{H}\left(\mathbf{x} \right)^T \mathbf{W} \mathbf{H}\left(\mathbf{x} \right)
\Delta \mathbf{x} = \mathbf{H}\left(\mathbf{x} \right)^T \mathbf{W} \left[ \mathbf{z} - \mathbf{h}\left(\mathbf{x} \right) \right]
\end{equation}
where $\mathbf{H}(\mathbf{x})$ denotes the Jacobian matrix of the measurement function with respect to the state vector $\mathbf{x}$, i.e.,
$\mathbf{H}(\mathbf{x}) = \frac{\partial \mathbf{h}(\mathbf{x})}{\partial \mathbf{x}}$.

However, iterative solving in \eqref{nonlinear2} and \eqref{nonlinear} relies heavily on the selection of initial voltage states, and is time-consuming for real-time monitoring of large-scale distribution grids. Meanwhile, the existing linearized methods inevitably introduce estimation errors in nonlinear distribution systems, especially across a broad range of operating conditions with DER integration. Furthermore, intrinsic to the WLS-based design, they rely on the Gaussian assumption and cannot handle unknown complex input statistics from DER generation \cite{NonGaussian8008784,NonGaussianvanin2023exact}.

\subsection{Gaussian Mixture Models for Uncertainty Representation}
In GMM, an unknown uncertainty distribution is represented as a weighted sum of multiple Gaussian components \cite{5298967GMM, hoeltgebaum2018generating}. 
It is expressed as a convex combination of $K$ Gaussian distributions:
\begin{equation}\label{GMM}
    y \sim \sum_{k=1}^K \pi_k \mathcal{N}\left(\mu_k, \sigma_k\right)
\end{equation}
where $\pi_k$ denotes the weight of the $k$th Gaussian component and satisfies $\sum_{k=1}^K \pi_k = 1$, 
$\mu_k$ and $\sigma_k$ are the mean vector and covariance matrix of the $k$th component, respectively \cite{5298967GMM}. 

In DSSE, the GMM model is employed to approximate unknown non-Gaussian uncertainties caused by DER generations, load variations, or
possible field errors in sensor measurements \cite{hoeltgebaum2018generating}. 
However, using expectation maximization to recursively solve the parameters in \eqref{GMM} is
computationally expensive, since each new data point $y_j$ multiplies the number of posterior terms by $K$. 
To avoid the computational burden, recent research \cite{ huang2019probabilistic} has developed data-driven probabilistic inference methods, such as mixture density network (MDN), for quantifying time-varying uncertainty. Nonetheless, without incorporating the intermittent characteristics of DER generation, the inferred GMM distributions may violate feasible generation limits and suffer from density leakage.

Therefore, this paper proposes a dual-branch PDIN for pseudo-measurement under unknown non-Gaussian uncertainty, which can mitigate density leakage beyond the physical boundaries of DER generation. The GMM parameters inferred by the PDIN are subsequently incorporated into the DSSE as detailed in the next section.

\begin{figure}[t]
  \centering
    \includegraphics[width=0.98\columnwidth]{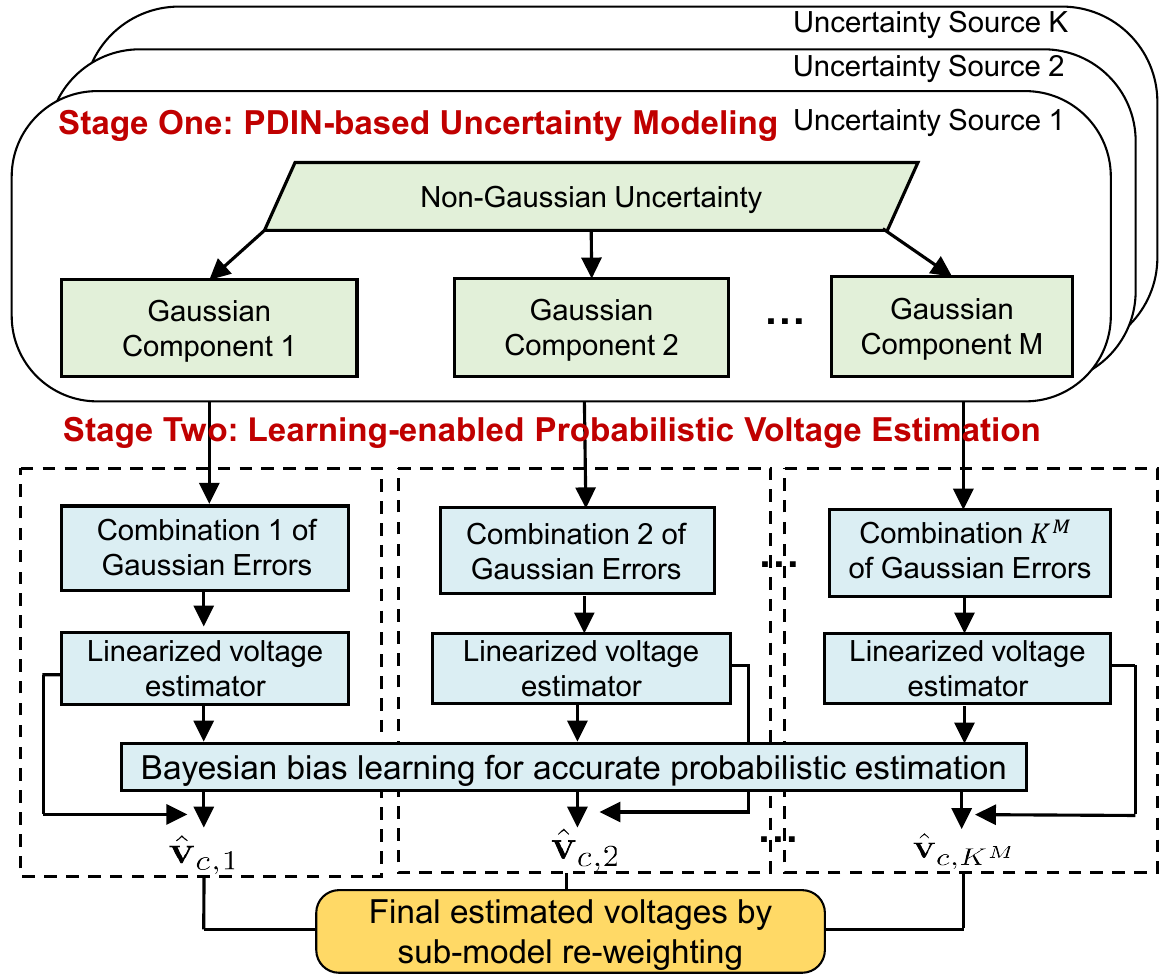}
    \caption{Proposed two-stage framework integrating probabilistic pseudo-measurement modeling into BBL-enabled state estimation.}
    \label{fig:proposed}
  \vspace{-10pt}
\end{figure}

\section{Proposed Two-Stage Learning-Enabled Probabilistic Voltage Estimation Algorithm}

This section proposes a BBL-enabled voltage estimation algorithm that integrates probabilistic pseudo-measurement modeling into a learning-enabled state estimation, as illustrated in Fig.~\ref{fig:proposed}. 
In Stage one, the proposed PDIN-based method learns the probability distribution of pseudo-measurements from historical data, parameterized as a Gaussian mixture that characterizes the arbitrary uncertainties. The resulting probabilistic components are subsequently fed to Stage two, where each Gaussian component combination serves as the input to a DSSE sub-model. The corresponding voltage estimation problems are solved in parallel by the proposed BBL-enabled linearized state estimator, enabling efficient voltage estimates under non-Gaussian uncertainties.

By explicitly quantifying and decomposing the unknown uncertainty into known Gaussian components in Section \ref{3-A} and developing a BBL-enabled surrogate of voltage estimation, the proposed two-stage algorithm is computationally efficient and accommodates arbitrary unknown DER uncertainties. 

\subsection{Stage One: Probabilistic Pseudo-Measurement Modeling under Unknown Uncertainty }\label{3-A}
A physics-constrained PDIN-based pseudo-measurement generation method is developed to jointly model DER pseudo-measurements and their probabilistic uncertainty. Unlike the existing single-task density learning methods, the proposed method accounts for the intermittency of DER generation by augmenting the modeling task into two output branches: a binary-state classification branch and a probability parameter estimation branch. Furthermore, the proposed methods develop statistics-informed activation and multi-task loss functions to enforce physical consistency.

\begin{figure}[!t]
\centering
     \includegraphics[width=0.47\textwidth]{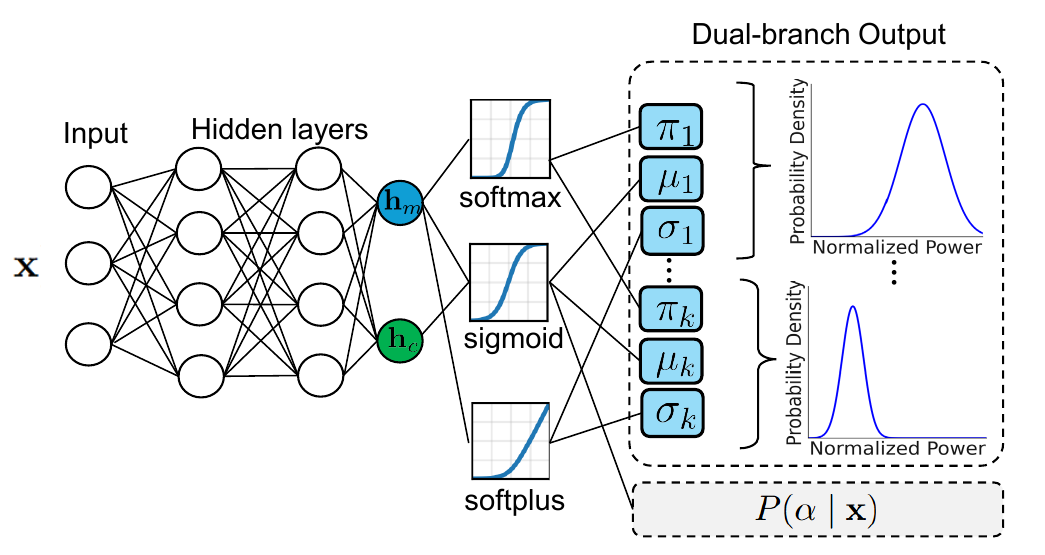} 
     \caption{Architecture of the proposed dual-branch probabilistic density inference network for non-Gaussian uncertainty quantification (Stage One).}
     \label{fig:PDIN-architecture}
\end{figure}

As shown in Fig.~\ref{fig:PDIN-architecture}, the proposed dual-branch PDIN architecture starts with a shared feature-extraction backbone followed by the two task-specific branches. Given an input as $\mathbf{x}\in \mathbb{R}^{d_1 \times d_2}$, $d_1$ denotes the number of previous time steps, and $d_2$ denotes the number of features consisting of the historical weather and power data. The latent representation $\mathbf{h}$ is first extracted through fully connected layers:
\begin{equation}
\mathbf{h}
=
\mathrm{ReLU}\!\left(
\mathbf{W}^{(L)} \cdots
\mathrm{ReLU}\!\left(
\mathbf{W}^{(1)} \mathbf{x} + \mathbf{b}^{(1)}
\right)
+ \mathbf{b}^{(L)}
\right)
\end{equation}
where $\mathbf{W}^{(l)}$ and $\mathbf b^{(l)}$ are the trainable weights and bias at layer $l$; $\mathrm{ReLU}(\cdot)$ denotes the rectified linear unit (ReLU) activation function.

The latent feature $\mathbf{h}$ is then projected into two branches, binary-state classification branch $\mathbf{h}_c$ and probability distribution estimation branch $\mathbf{h}_m$:
\begin{equation}
    \mathbf{h}_c=\mathrm{ReLU}(\mathbf{W}_c \mathbf h+\mathbf b_c)
\end{equation}
\begin{equation}
\quad \mathbf{h}_m=\mathrm{ReLU}(\mathbf{W}_m \mathbf h+\mathbf b_m)
\end{equation}
where $\mathbf{W}_{c}$, $\mathbf{W}_{m}$, $\mathbf b_c$, and $\mathbf b_m$ are the trainable parameters. 

The binary-state classification branch estimates the conditional probability $P(\alpha\mid \mathbf{x})$ for generation-state identification. This branch aims to predict the intermittency of renewable generation and avoid a distribution covering negative values. The probability of the binary generation states on this branch is then obtained as:
\begin{equation}\label{z}
P(\alpha \mid \mathbf{x}) = \mathrm{sigmoid}\!\left(\mathbf{W}_c \, \boldsymbol{\phi}_c(\mathbf{h}_c) + \mathbf b_c\right)
\end{equation}
where $\alpha\in\{0,1\}$ denotes the generation-state indicator; $\boldsymbol{\phi}_c(\cdot)$ denotes the branch-specific hidden transformation; the sigmoid function is used as the activation function to confine the classification probability to [0, 1]. 

The \textit{probability distribution parameter branch} $\mathbf{h}_m$ is used to estimate the GMM parameters to characterize the distribution of the normalized power output $y$, conditioned on the unit being in the generation state:
\begin{equation}\label{dist}
p(y \mid \mathbf{x}) = \sum_{k=1}^K \pi_k(\mathbf{x}) \, \mathcal{N}\!\big(y \mid \mu_k(\mathbf{x}), \sigma_k^2(\mathbf{x})\big)
\end{equation}
where $\mathcal{N}(\cdot)$ denotes the Gaussian probability density function.

 The mixture parameters of the GMM are then obtained through the selective activation functions to enforce the physical consistency:
\begin{align}
\boldsymbol{\pi}(\mathbf{x})   &= \mathrm{softmax}\!\left( \mathbf{W}_{\pi}\,\boldsymbol{\phi}_m(\mathbf{h}_m) + \mathbf{b}_{\pi} \right) \label{eq:pi}\\
\boldsymbol{\mu}(\mathbf{x})   &= \mathrm{sigmoid}\!\left( \mathbf{W}_{\mu}\,\boldsymbol{\phi}_m(\mathbf{h}_m) + \mathbf{b}_{\mu} \right) \label{eq:mu}\\
\boldsymbol{\sigma}(\mathbf{x})&= \mathrm{softplus}\!\left( \mathbf{W}_{\sigma}\,\boldsymbol{\phi}_m(\mathbf{h}_m) + \mathbf{b}_{\sigma} \right) \label{eq:sigma}
\end{align}
where $\mathbf{W}_{\pi}, \mathbf{b}_{\pi}, \mathbf{W}_{\mu}, \mathbf{b}_{\mu}, \mathbf{W}_{\sigma}$, and $\mathbf{b}_{\sigma}$ are the weights and biases for the GMM representation; $\boldsymbol{\phi}_m(\cdot)$ denotes the hidden transformation for $\mathbf{h}_m$.

Note that the activation functions in~\eqref{eq:pi}–\eqref{eq:sigma} are designated to enforce the statistical significance of $\boldsymbol{\pi}$, $\boldsymbol{\mu}$, and $\boldsymbol{\sigma}$. The softmax activation in~\eqref{eq:pi} produces normalized mixture weights, with the $k$-th component weight given by:
\begin{equation}
\pi_k(\mathbf{x})
= \frac{\exp(\mathbf{w}_{\pi,k}^{\top}\mathbf h_m + b_{\pi,k})}
       {\sum_{j=1}^K \exp(\mathbf{w}_{\pi,j}^{\top}\mathbf h_m + b_{\pi,j})},
\quad k=1,\ldots,K
\end{equation}
which guarantees $\pi_k(\mathbf{x}) > 0$ and $\sum_{k=1}^K \pi_k(\mathbf{x})=1$, thus forming a valid probability distribution. 

The sigmoid activation in~\eqref{eq:mu} maps real-valued inputs into the unit interval, guaranteeing that the component means are bounded between 0 and 1, while the softplus activation in~\eqref{eq:sigma} enforces the positivity of the standard deviations.

The proposed dual-branch PDIN-based method is trained by minimizing the mean composite loss over the training dataset of size $N$, which jointly accounts for generation-state identification and physics-bounded probabilistic estimation. The loss function is formulated as:
\begin{equation}
\mathcal{L}_{\mathrm{total}}
=
\frac{1}{N}
\sum_{i=1}^{N}
\big[
\mathcal L_{\mathrm{class},i}
+
\alpha_i \left(
\mathcal L_{\mathrm{nll},i}
+
\lambda \mathcal L_{\mathrm{b},i}
\right)
\big]
\end{equation}
where $\lambda$ denotes the  penalty coefficient;
$\mathcal L_{\mathrm{class},i}$ denotes the standard binary cross-entropy for the classification:
\begin{equation}
\mathcal L_{\mathrm{class},i}
=
-
\big[
\alpha_i \log P(\alpha_i\mid \mathbf{x}_i)
+
(1-\alpha_i)\log(1-P(\alpha_i\mid \mathbf{x}_i))
\big]
\end{equation}

The distribution loss is given by the negative log-likelihood of the predicted GMM:
\begin{equation}
\mathcal L_{\mathrm{nll},i}
=
-
\log
\left(
\sum_{k=1}^{K}
\pi_k(\mathbf{x}_i)
\mathcal{N}\!\left(
y_i \mid \mu_k(\mathbf{x}_i), \sigma_k^2(\mathbf{x}_i)
\right)
\right)
\end{equation}

The boundary penalty term $\mathcal L_{\mathrm{b},i}$ penalizes probability mass outside the normalized output domain $[0,1]$, calculated by:
\begin{equation}
\begin{aligned}
\mathcal L_{\mathrm{b},i}
&=
\sum_{k=1}^{K}
\pi_k(\mathbf{x}_i)
\Bigg[
\Phi\!\left(
\frac{0-\mu_k(\mathbf{x}_i)}{\sigma_k(\mathbf{x}_i)}
\right)
\\
&\qquad\quad
+
1-
\Phi\!\left(
\frac{1-\mu_k(\mathbf{x}_i)}{\sigma_k(\mathbf{x}_i)}
\right)
\Bigg]
\end{aligned}
\end{equation}
where \(\Phi(\cdot)\) denotes the cumulative distribution function of the standard normal distribution. This addresses the aforementioned density-leakage issue inherent in conventional probabilistic inference.

During the inference, for any new input $\mathbf{x}$, the weel-trained PDIN method produces generation state probability and distribution estimation by \eqref{eq:pi}-\eqref{eq:sigma}, parameterized by $\left\{\pi_k(\mathbf{x}), \mu_k(\mathbf{x}), \sigma_k(\mathbf{x})\}_{k=1}^K \right\}$.
Unlike most pseudo-measurement generation methods that yield only point estimates with an implicit error distribution, the proposed method infers the GMM-represented probabilistic distribution of DER generation. Next, this resulting probabilistic representation will be embedded in the proposed learning-enabled linearized voltage estimation formulation.
\vspace{-5pt}
\subsection{Bridging Stages One \& Two: Proposed Learning-Enabled Linearized Estimation Framework} \label{3-B}
The proposed uncertainty modeling on Stage One explicitly characterizes the error behavior of DER inputs in a probabilistic GMM form. 
Given $M$ unknown uncertainty sources across the grid, each modeled by $K$ Gaussian components, the density parameters obtained from Stage~One are expressed as:
\begin{equation}
\{\pi^{i}_k,\, \mu^{i}_k,\, \sigma^{i}_k\}, 
\quad i = 1,\dots,M~\text{and}~ k = 1,\dots,K
\label{eq:gmm_params}
\end{equation}
where $\pi^i_k$, $\mu^i_k$, and $\sigma^i_k$ denote the weight, mean, and standard deviation of 
the $k$-th Gaussian component in the $i$-th non-Gaussian uncertainty source. These parameters form a total of $K^M$ combinations of the Gaussian-component combination, denoted as $\mathcal{C}$.
%{c_{ik}}

For each combination $c \in \mathcal{C}$, the measurement vector in the $c$-th state estimation sub-model is expressed by:
\begin{equation} \mathbf{z}_{c} = \big[\, \mathbf{z}_g^{\top},\,  \boldsymbol{\mu}_{\mathrm{c}}^{\top}\,\big]^{\top} \label{eq:z_c} \end{equation}
where $\mathbf{z}_{g}$ denotes the measurement vector originally obeying the Gaussian assumption; $\boldsymbol{\mu}_{\mathrm{c}}=\{\mu_k^i\}_{i=1}^M$ denotes the vector of mean values of the pseudo-measurements resulting from \eqref{eq:gmm_params}.

For each Gaussian-component combination $c$, the corresponding standard 
deviations obtained from Stage One are used to characterize the uncertainty 
of the pseudo-measurements. Specifically, the measurement covariance matrix 
associated with combination $c$ is constructed as:
\begin{equation}
\mathbf{R}_c =
\begin{bmatrix}
\mathbf{R}_g & \mathbf{0}\\
\mathbf{0} & \mathbf{\Sigma}_c
\end{bmatrix}
\text{ with }
\mathbf{\Sigma}_c =
\mathrm{diag}\left(
(\sigma^1_{c})^2,\ldots,
(\sigma^M_{c})^2
\right)
\label{Covar}
\end{equation}
where $\mathbf{R}_g$ denotes the covariance matrix of the original 
Gaussian measurements and $\mathbf{\Sigma}_c$ characterizes the uncertainty 
of the pseudo-measurements in combination $c$. Accordingly, the 
combination-specific weighting matrix is given by 
$\mathbf{W}=\mathbf{R}_c^{-1}$.

The parameterization in \eqref{eq:gmm_params} enables the use of the standard DSSE formulation under the Gaussian assumption, as in \eqref{se}-\eqref{nonlinear}, which is nonlinear. However, given multiple non-Gaussian uncertainty sources arising from growing DER installations or non-conventional field noise, repeatedly solving the nonlinear DSSE for each $\mathbf{z}_{c}$ becomes computationally prohibitive. 
To achieve the accuracy and computational efficiency, a learning-enabled linearized DSSE approach is proposed as a data-aided analytical surrogate, providing fast yet probabilistic voltage results. 
\begin{figure}[!t]
\centering
     \includegraphics[width=0.5\textwidth]{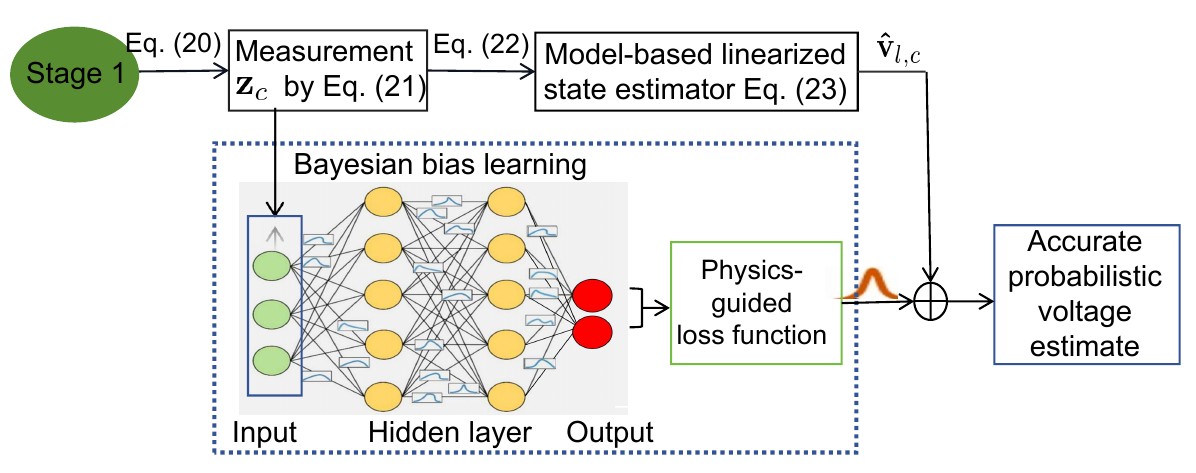} 
     \caption{Architecture of the proposed physics-guided Bayesian bias learning network for 
accurate probabilistic voltage estimation (Stage Two).}
     \label{fig:brn-architecture}
\end{figure}
The approximated linear solution of voltage states according to $\mathbf{z}_{c}$ 
is estimated by the following linearized estimator:
\begin{equation}
\hat{\mathbf{v}}_{l,c}
= ({\mathbf{H}_l}^{\top}\mathbf{W}\mathbf{H}_l)^{-1}
  {\mathbf{H}_l}^{\top}\mathbf{W}\mathbf{z}_{c}
\label{eq:linear_estimator}
\end{equation}
where $\hat{\mathbf{v}}_{l,c}$ denotes the linear estimation solution;
$\mathbf{H}_l$ denotes the constant Jacobian matrix based on the distribution network parameters, which is derived from a set of linearized measurement functions for available sensors or pseudo-measurements in the distribution system. The construction of $\mathbf{H}_l$ is briefly described in the Appendix due to the page limit.

As shown in Fig. \ref{fig:brn-architecture}, the BBL-based probabilistic voltage estimation algorithm is proposed to compensate for the linearization errors introduced by \eqref{eq:linear_estimator}.
This Bayesian approach achieves uncertainty-aware bias learning and improves generalization with limited training data by embedding physics knowledge.
The accurate voltage estimate based on $\mathbf{z}_{c}$ is parameterized as a sum of a linearized solution \eqref{eq:linear_estimator} and learnable linearization biases as:
\begin{equation}
\hat{\mathbf{v}}_c = \mathbf{\hat v}_{l,c}+\Delta\mathbf{v}
\label{eq:bnn_target_error}
\end{equation}
where $\Delta\mathbf{v}=g_{\boldsymbol{\omega}}(\mathbf{z}_c,\mathbf{\hat v}_{l,c})$ denotes the linearization error vector. We intend to estimate $\Delta\mathbf{v}$ by the proposed BBL-based neural network, which is detailed in the next subsection.

\subsection{Stage Two: Bayesian Bias Learning for Voltage Estimation} \label{3-B}
 The proposed BBL-based method parameterizes the nonlinear DSSE formulation as a combination of the linear voltage estimation solution \eqref{eq:linear_estimator} and its probabilistic biases induced by the linearization. The probabilistic distribution of the voltage bias is effectively estimated in a Bayesian learning framework, which models the network weights and bias $\boldsymbol{\omega}=\{\mathbf{W}_\ell,\mathbf{b}_\ell\}_{\ell=1}^L$ as random variables following learnable probability distributions, rather than deterministic values. Original Bayesian learning often favors Gaussian priors, which require learning only two parameters per neural network weight to describe the statistical distribution. However, selecting a complex distribution type with more parameters may cause the curse of dimensionality. By inferring uncertainty and decomposing into Gaussian components in Stage One, our BBL-based method, as an enhanced Bayesian learning approach, effectively mitigates this issue.

 The proposed BBL model $g_{\boldsymbol{\omega}}(\cdot)$ is constructed to offline learn the mapping from measurements to the bias between linearized and nonlinear state estimates, thereby adaptively compensating for the voltage bias introduced by linearization during online inference. Given a dataset with $N$ samples $\mathcal{D}=\{(\mathbf{z}^{(s)},\,\mathbf{v}^{(s)})\}_{s=1}^{N}$, $\mathbf{v}^{(s)}$ denotes the accurate voltage estimates. 
 The linear solution $\mathbf{\hat v}_l^{(s)}$ are solved by \eqref{eq:linear_estimator}, and the bias $\Delta\mathbf{v}^{(s)}=\mathbf{\hat v}^{(s)}-\mathbf{\hat v}_l^{(s)}$ are collected for the BBL training.
  The bias correction is modeled as an $L$-layer Bayesian neural network, with a bias correction $\Delta\mathbf{v}$ as the output:
\begin{equation}
\Delta\mathbf{v}
=g_{\boldsymbol{\omega}}(\mathbf{u})
+\boldsymbol{\varepsilon}
\label{eq:bnn_likelihood}
\end{equation}
where $\mathbf{u}=
\begin{bmatrix}\mathbf{z}^\top &
\hat{\mathbf{v}}_{l}^\top
\end{bmatrix}^\top$ denotes the input vector;
% ;
$\boldsymbol{\varepsilon}$ represents the observation noise.

Within a Bayesian formulation, a prior distribution is imposed on the network parameters $\boldsymbol{\omega}$ to regularize the model and encode prior belief before observing data.
Here, a zero-mean Gaussian prior $p(\boldsymbol{\omega})
=\mathcal{N}\!\left(\mathbf{0},\,\alpha^{2}\mathbf{I}\right)$ is adopted  \cite{blundell2015weight}, where $\alpha$ controls the prior precision and $\mathbf{I}$ is the identity matrix.
Given the training dataset $\mathcal{D}$, the posterior distribution of the BBL weights is used to model the conditional distribution of the bias correction given the input features and network weights, according to Bayes' rule as:
\begin{equation}
\begin{aligned}
p(\boldsymbol{\omega}\mid\mathcal{D})
&=
\frac{
p(\mathcal{D}\mid\boldsymbol{\omega})\,p(\boldsymbol{\omega})
}{
p(\mathcal{D})
} \\
&\propto
p(\boldsymbol{\omega})
\prod_{k=1}^N
\mathcal{N}\!\left(
\Delta\mathbf{v}^{(k)}
;\,
g_{\boldsymbol{\omega}}(\mathbf{u}^{(k)}),
\,\sigma^2\mathbf{I}_n
\right)
\end{aligned}
\label{eq:bnn_posterior}
\end{equation}

Due to the nonlinear structure of $g_{\boldsymbol{\omega}}(\cdot)$, the posterior distribution $p(\boldsymbol{\omega}\mid\mathcal{D})$ is analytically intractable. 
Hence, variational inference is adopted to approximate the posterior by a tractable distribution $q(\boldsymbol{\omega})$.  
In Bayesian learning, the optimal variational distribution is obtained by minimizing the Kullback-Leibler (KL) divergence:
\begin{equation}
q^{*}(\boldsymbol{\omega})
=
\arg\min_{q(\boldsymbol{\omega}) \in \mathcal{Q}}
\mathrm{KL}\!\left(
q(\boldsymbol{\omega})
\;\|\;
p(\boldsymbol{\omega}\mid\mathcal{D})
\right)
\label{eq:vi_kl}
\end{equation}
which $q(\boldsymbol{\omega})$ is initialized as a predefined Gaussian distribution type $
\mathcal{N}\!\left(\boldsymbol{\mu}_{\boldsymbol{\omega}},\,\mathrm{diag}(\boldsymbol{\sigma}_{\boldsymbol{\omega}}^2)\right)$, and $\boldsymbol{\mu}_{\boldsymbol{\omega}}$ and $\boldsymbol{\sigma}_{\boldsymbol{\omega}}$
represent the mean and standard deviation, respectively. The objective \eqref{eq:vi_kl} is equivalent to maximizing the evidence lower bound $\mathbb{E}_{q}(\boldsymbol{\omega})
\!\left[
\log p(\mathcal{D}\mid\boldsymbol{\omega})
\right]
-
\mathrm{KL}\!\left(
q_{\boldsymbol{\phi}}(\boldsymbol{\omega})
\;\|\;
p(\boldsymbol{\omega})
\right)$.

Substitute \eqref{eq:bnn_posterior} to \eqref{eq:vi_kl}, and the proposed BBL model $g_{\boldsymbol{\omega}}(\cdot)$ is trained to update $\{\boldsymbol{\mu}_{\boldsymbol{\omega}},\boldsymbol{\sigma}_{\boldsymbol{\omega}}\}$ by minimizing the following composite loss function, which consists of an expected bias fitting term and Bayesian regularization:
\begin{equation}
\begin{aligned}
\mathcal{L}_{\mathrm{bbl}}
=\frac{1}{N}
\sum_{k=1}^N
\mathbb{E}_{\boldsymbol{\omega}\sim q}
\!\Big[
\big\|
g_{\boldsymbol{\omega}}(\mathbf{u}^{(k)})
- \Delta\mathbf{v}^{(k)}
\big\|_2^2
\Big]
\\
\quad+ \beta\,\mathrm{KL}\!\left(q(\boldsymbol{\omega})
\,\big\|\,p(\boldsymbol{\omega})
\right)
\end{aligned}
\label{eq:bnn_loss_final}
\end{equation}
where the first term measures the squared error between the predicted bias and the true bias;
$\mathrm{KL}\!\left(
q(\boldsymbol{\omega})
\,\big\|\,
p(\boldsymbol{\omega})
\right)$ regularizes the variational posterior, preventing overfitting and facilitating the uncertainty quantification.
\begin{algorithm}[!t]
\caption{Proposed Bayesian Bias Learning (Stage Two)}\label{Offline}
\label{alg:brn_stage_two}
% \small
\textbf{Offline Training Phase}
\begin{algorithmic}[1]
\STATE \textbf{Input:} Training dataset $\mathcal{D}=\{(\mathbf{u}^{(k)},\,\Delta\mathbf{v}^{(k)})\}_{k=1}^{N}$, prior distribution of network weights, learning rate $\eta$
\STATE Initialize posterior distribution $q(\boldsymbol{\omega})$ 
\FOR{each training epoch}
    \STATE Sample the latest network parameters $\boldsymbol{\omega} \sim q(\boldsymbol{\omega})$
    \STATE Optimal posterior distribution by KL divergence \eqref{eq:vi_kl}
    \STATE Update the parameters of the posterior distribution $q(\boldsymbol{\omega})$ via gradient descent based on \eqref{eq:bnn_loss_final}:\\
$\boldsymbol{\mu}_{\boldsymbol{\omega}}
\leftarrow
\boldsymbol{\mu}_{\boldsymbol{\omega}}
-
\eta\,
\nabla_{\boldsymbol{\mu}_{\boldsymbol{\omega}}}
\mathcal{L}_{\mathrm{bbl}}$,
$\boldsymbol{\sigma}_{\boldsymbol{\omega}}
\leftarrow
\boldsymbol{\sigma}_{\boldsymbol{\omega}}
-
\eta\,
\nabla_{\boldsymbol{\sigma}_{\boldsymbol{\omega}}}
\mathcal{L}_{\mathrm{bbl}}$
\ENDFOR
\STATE \textbf{Output:} Trained BBL model $g_{\boldsymbol{\omega}}(\cdot)$ with its posterior $q(\boldsymbol{\omega})$
\end{algorithmic}
\end{algorithm}
The pseudocode of the proposed DSSE algorithm is summarized in Algorithm~\ref{Offline} for the offline training. Once the BBL model $g_{\boldsymbol{\omega}}(\cdot)$ is trained offline via stochastic gradient descent, the distribution of the bias correction for a new input $\mathbf{u}$ is characterized by the posterior
$q(\boldsymbol{\omega})$ as:
\begin{equation}
p\!\left(
\Delta\mathbf{v}_d
\mid \mathbf{u}
\right)
=
\int
\mathcal{N}\!\left(
\Delta\mathbf{v}_d;\,
g_{\boldsymbol{\omega}}(\mathbf{u}),\,
\sigma^2\mathbf{I}_n
\right)
q(\boldsymbol{\omega})
\,\mathrm{d}\boldsymbol{\omega}
\label{eq:bnn_predictive}
\end{equation}
where $\Delta\mathbf{v}_d$ denotes the random bias correction vector following the posterior distribution. 
This formulation represents a Bayesian model average over the variational posterior of the network parameters.

The posterior mean of \eqref{eq:bnn_predictive} is obtained by \cite{heckerman2022tutoriallearningbayesiannetworks}:
\begin{equation}
\Delta\mathbf{v}
=
\mathbb{E}\!\left[\Delta\mathbf{v}_d \mid \mathbf{u}\right]
=
\frac{1}{T}
\sum_{t=1}^{T}
g_{\boldsymbol{\omega}^{(t)}}(\mathbf{u}),
\qquad
\boldsymbol{\omega}^{(t)} \sim q(\boldsymbol{\omega})
\label{eq:bnn_mc_mean}
\end{equation}
where $T$ denotes the number of samples drawn from the variational posterior, and $\boldsymbol{\omega}^{(t)}$ represents the
$t$-th sampled network parameter realization.

The corresponding covariance, which characterizes the uncertainty associated with the learned residual correction, is computed as:
\begin{equation}
\begin{aligned}
\boldsymbol{\Sigma}_{\mathrm{bbl}}
&=
\frac{1}{T-1}
\sum_{t=1}^{T}
\Big(
g_{\boldsymbol{\omega}^{(t)}}(\mathbf{u})
-
\mathbb{E}\!\left[\Delta\mathbf{v}_d \mid \mathbf{u}\right]
\Big)
\\
&\quad \times
\Big(
g_{\boldsymbol{\omega}^{(t)}}(\mathbf{u})
-
\mathbb{E}\!\left[\Delta\mathbf{v}_d \mid \mathbf{u}\right]
\Big)^{\top}
\end{aligned}
\label{eq:bnn_mc_cov}
\end{equation}

Therefore, the proposed algorithm effectively produces a probabilistic distribution of the estimated voltages, parameterized by \eqref{eq:bnn_mc_mean} and \eqref{eq:bnn_mc_cov}.

\subsection{Online Voltage Estimation and Weighted Model Fusion}
The pseudocode of online voltage estimation under non-Gaussian uncertainty is shown in Algorithm~\ref{alg:brn_online}.
During online implementation, each measurement vector $\mathbf{z}_c$ inferred in Stage One is fed into the trained BBL model to estimate accurate nodal voltages in parallel.
The trained BBL-based method serves as an effective estimator for various measurement vectors due to its generalizability and physics-informed nature, and is directly applied to each $\mathbf{z}_c$ associated with a Gaussian-component combination.

For each combination $c \in \mathcal{C}$, the accurate state estimate $\hat{\mathbf{v}}_{c}$ is obtained by combining the proposed bias learning and an analytic linear solution. The point estimate of this probabilistic state estimate is obtained as:
\begin{equation}
\hat{\mathbf{v}}_{c}
=\mathbf{\hat{v}}_{l,c}+g_{\boldsymbol{\omega}}(\mathbf u)
\label{eq:bnn_final_estimate}
\end{equation}

\begin{algorithm}[!t]
\caption{Online Probablistic Voltage Estimation}
\label{alg:brn_online}
\begin{algorithmic}[1]
\STATE \textbf{Input:} Measurement vector $\mathbf{z}_c$ obtained by Stage One, well-trained $g_{\boldsymbol{\omega}}(\cdot)$ in Stage Two
\STATE Compute voltage estimates $\hat{\mathbf{v}}_l$ using the linearized DSSE formulation \eqref{eq:linear_estimator} to form
 BBL input $\mathbf{u} = [\mathbf{z}^\top,\,\hat{\mathbf{v}}_l^\top]^\top$
\FOR{each $c \in \mathcal{C}$}
    \STATE Sample $\mathbf{\omega}^{(t)} \sim q(\mathbf{\omega})$, and compute bias correction by the well-trained ${\Delta\mathbf{v}} = g_{\mathbf{\omega}}(\mathbf{u})$
    \STATE Compute the mean bias correction $\mathbb{E}\!\left[\Delta\mathbf{v}\mid \mathbf{u}\right]$ by \eqref{eq:bnn_mc_mean} and covariance $\mathbf{\Sigma}_{\mathrm{bbl}}$ by \eqref{eq:bnn_mc_cov}
    \STATE Obtain BBL-corrected state estimate by \eqref{eq:bnn_final_estimate}, including state estimate $\hat{\mathbf{v}}_{c}$ and covariance  $\mathbf{\Sigma}_{\mathrm{bbl}}$
\ENDFOR
\STATE \textbf{Output:} Final voltage estimate $\hat{\mathbf{v}}$ by \eqref{eq:x_final} and variance of the final estimated voltages $\mathbf{\Sigma}_{\hat{\mathbf{v}}}$ by \eqref{eq:final_state_variance}
\end{algorithmic}
\end{algorithm}

After obtaining the estimated states $\hat{\mathbf{v}}_{c}$ for all Gaussian-component combinations, the final system state is computed as the weighted sum of all the sub-model estimates:
\begin{equation}
\hat{\mathbf{v}} = 
\sum_{c \in \mathcal{C}} w_c \, \hat{\mathbf{v}}_{c} 
\quad \text{with} \quad 
w_{c} = \prod_{i=1}^{M} \pi_{i,c(i)},
\label{eq:x_final}
\end{equation}
where $w_c$ denotes the normalized weight of the $c$-th combination and satisfies $\sum_{c \in \mathcal{C}} w_c = 1$; the combination weight $w_c$ for the DSSE submodel fusion is determined by the product of $\pi_{i,c(i)}$, which is estimated by \eqref{eq:pi} in Stage One.

Furthermore, the variance of the final estimated voltages is determined as:
\begin{equation}
\mathbf{\Sigma}_{\hat{\mathbf{v}}}
=
\sum_{c \in \mathcal{C}}
w_c
\left(\mathbf{\Sigma}_\mathrm{bbl}
+\left\|
\hat{\mathbf{v}}_c - \hat{\mathbf{v}}
\right\|^2
\right)
\label{eq:final_state_variance}
\end{equation}

Overall, the proposed BBL-based voltage estimation algorithm combines an analytical linearized solution with learning-enabled bias offset, achieving high estimation accuracy under unknown uncertainties.

\textbf{Remarks}: The uncertainty parameters estimated in Stage One can be periodically updated by offline training to capture evolving patterns, thus providing flexibility. Our results indicate that the offline training for the uncertainty modeling takes only minutes, enabling timely refreshes that support hour-ahead or near-real-time model updates for sequential system-wide voltage estimation.

\section{Case Study}
The proposed algorithm is tested on the unbalanced IEEE 13- and 123-bus distribution feeders \cite{Feeder}. The 13-bus system is modified with two PV units and one wind turbine generator (WTG), while six DERs are installed in the 123-bus system.
 The following conditions are applied to the maximum errors of the hybrid metering infrastructure by setting 0.7\% for voltage phasors in PMU data, 2\% for active and reactive power flows in SCADA data, and 20\% for power measurements (pseudo-measurements are used here as a practical consideration \cite{manitsas2012distribution, mestav2019bayesian}).  
The installation locations of these sensors in these test systems are shown in Table \ref{tab:measurement_placement}. Low PMU coverage is set, with 2 PMUs installed in the 13-bus system and 6 PMUs installed in the 123-bus system \cite{zhangInterval}.
To investigate the impact of DER-induced uncertainties on voltage estimation, two raw DER generation datasets \cite{OEDI_Dataset_4568,winddata} are used, and our analysis indicates that the DER uncertainty exhibits complex non-Gaussian characteristics. They are considered on bus~9 in the 13-bus system and on buses~14, 61, and~151 in the 123-bus system.
\begin{table}[!t]
\centering
\captionsetup{font=small,justification=centering}
\caption{\textsc{Measurement Placement of Sensors in Test Distribution Systems}}
\begin{tabular}{|c|c|c|}
\hline
\textbf{Sensor Types} & \textbf{13-bus System} & \textbf{123-bus System} \\ \hline
SCADA & \makecell{2-3, 4-5,\\ 11-7} & 
\makecell{1-7, 9-14, 15-16, 13-52, \\ 18-35, 44-45,
57-60, 76-77, \\ 86-87, 99-100,
110-112} \\ \hline
PMU & \makecell{6, 8} & \makecell{149, 8, 25,\\ 54, 97, 108} \\ \hline
\end{tabular}
\label{tab:measurement_placement}
\vspace{-6pt}
\end{table}
To evaluate voltage estimation performance, Monte Carlo simulations are run to randomly sample measurements with uncertain errors.
The generated measurement samples (2000 and 8000 for the 13- and 123-bus systems, respectively) are then divided into 70\% for training, 
10\% for validation, and 20\% for testing. 

For DERs with distinct uncertainty characteristics, we train separate PDIN models to capture their source-specific probability distributions. 
Each PDIN takes the historical power and meteorological features over 
the previous 24 time steps as input. The shared feature-extraction 
network consists of two fully connected hidden layers with 128 and 
64 neurons, respectively, followed by two task-specific branches with 
32 hidden neurons. ReLU is used in the hidden layers, while sigmoid 
is adopted for the generation-state probability and GMM component 
means, softmax for the GMM component weights, and softplus for the 
component standard deviations. The number of Gaussian components is 
set to $K=3$. Each PDIN is trained using the Adam optimizer with a 
learning rate of $10^{-4}$ and a batch size of 64 for a maximum of 
300 epochs. Early stopping with a patience of 20 epochs is applied 
according to the validation loss.

The BBL model consists of two Bayesian fully connected hidden layers 
with 128 and 64 neurons with ReLU activation. The input 
to the BBL model consists of the measurement vector and the corresponding 
linearized voltage estimate, as defined in \eqref{eq:bnn_target_error}, and its output is the 
linearization-induced voltage bias. A zero-mean unit-variance Gaussian 
prior $\mathcal{N}(\mathbf{0},\mathbf{I})$ is assigned to the network 
parameters, while the variational posterior is represented by a 
diagonal Gaussian distribution. The BBL model is trained using the 
Adam optimizer with a learning rate of $10^{-3}$ and a batch size of 
64 for a maximum of 300 epochs. The coefficient of the KL-divergence 
regularization term in \eqref{eq:vi_kl} is set to $\beta=10^{-4}$. During online 
inference, $T=50$ network parameter samples are drawn from the 
variational posterior to calculate the posterior mean and covariance 
of the bias correction by \eqref{eq:bnn_predictive} and \eqref{eq:bnn_mc_mean}, respectively.

 The mean absolute error (MAE), the root mean square error (RMSE), and the maximum error are used to evaluate the accuracy of voltage estimation. Furthermore, probabilistic metrics including continuous ranked probability score (CRPS), prediction interval coverage probability (PICP), and mean prediction interval width (MPIW) \cite{fu2027polar} are adopted, where the nominal confidence level is 95\%. The following three recent DSSE methods, including the recent two-stage estimators, are compared:
\begin{itemize}
\item \textbf{WLS w/DGA}:
The Gaussian component with the largest GMM weight is chosen to form a single-Gaussian approximation of the pseudo-measurement uncertainty, i.e., dominant Gaussian approximation (DGA). The Gaussian distribution parameters are then used in a WLS-based nonlinear estimator \cite{liu2019} to estimate the voltages.
\item \textbf{ANN-WLS}\cite{manitsas2012distribution}: This two-stage method first models the pseudo-measurements and their non-Gaussian errors by ANNs in the GMM form. Then, the nonlinear WLS-based estimator for distribution systems is used. 
\item \textbf{DBN-WLS} \cite{huang2019probabilistic}:
The pseudo-measurements are generated using DBNs, and their associated non-Gaussian errors are characterized by GMMs. Similar to ANN-WLS and ours, this method is two-stage, and the voltage estimation adopts a physical model-based WLS estimator.

% \item \textcolor{blue}{Direct DNN-based DSSE \cite{azimian2022state}:} No-Gassian
\end{itemize}

Furthermore, to evaluate the value of the proposed physics-guided design in Stage Two, Section IV-D provides an ablation study by replacing the proposed physics-guided BBL-based method with a residual neural network (ResNN), while preserving Stage One.
\vspace{-5pt}
\subsection{Uncertainty Quantification of Pseudo-Measurement}
The effectiveness of the proposed PDIN-based uncertainty modeling approach is evaluated and compared with other data-driven methods. The adopted DER generation dataset contains records of power outputs and meteorological features 
at a 15-minute resolution. 
A total of 30,000 samples 
are used for training, validation, and testing.
\begin{table}[!t]
\centering
\captionsetup{font=small,justification=centering}
\caption{\textsc{Performance Comparison for Non-Gaussian Uncertainty Quantification}}
\small
\begin{tabular}{|c|c|c|c|
}
\hline
Metric & ANN \cite{manitsas2012distribution}& DBN \cite{huang2019probabilistic} & \textbf{Proposed PDIN} \\
\hline
MAE [p.u.]  & 0.0514    & 0.0278    & \textbf{0.0119} \\
\hline
RMSE [p.u.] &  0.0690   & 0.0366    & \textbf{0.0232} \\
\hline
\end{tabular}
\label{tab:pdin}
\end{table}

\begin{figure}[!t]
    \centering
        \includegraphics[width=0.46\textwidth]{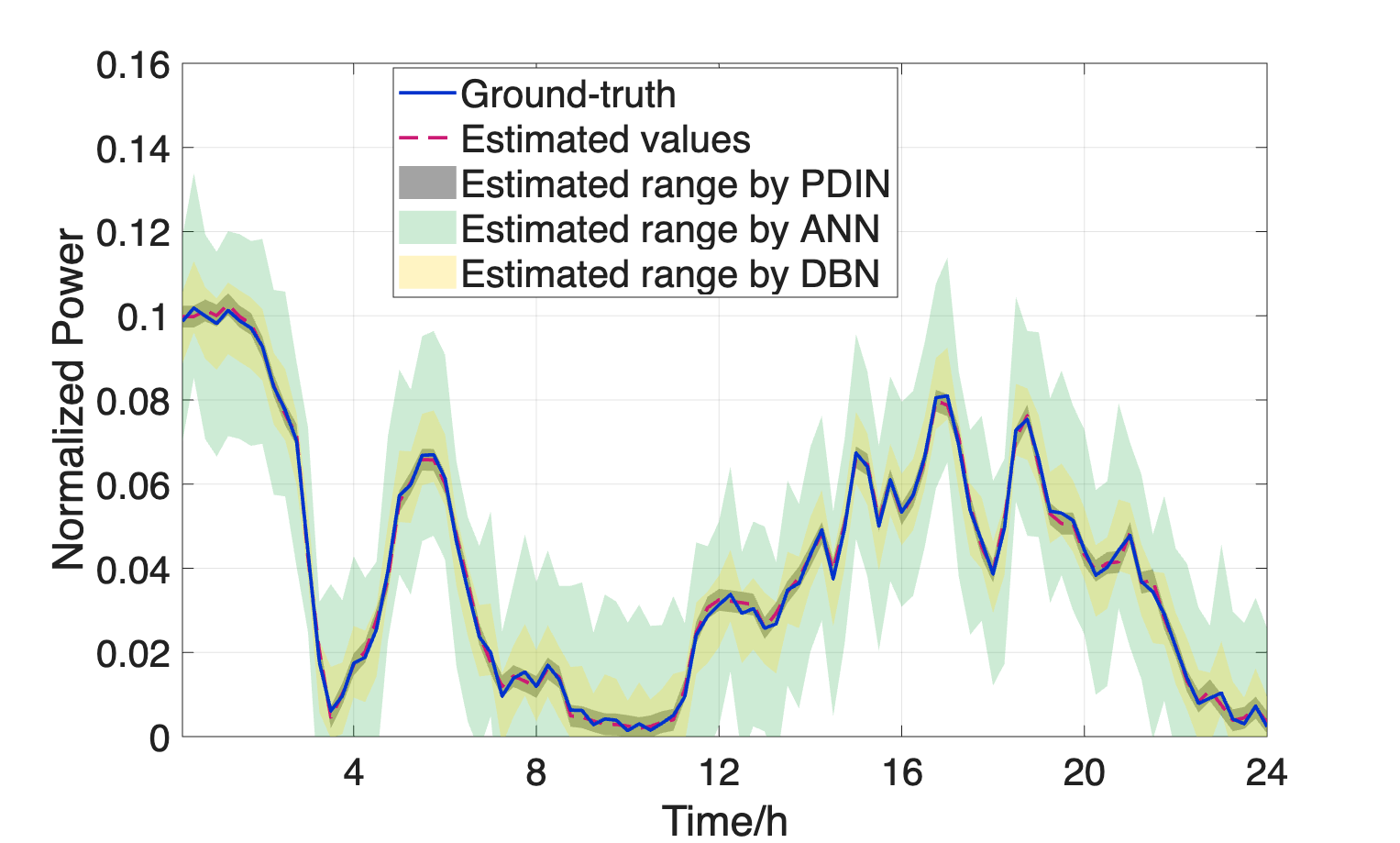}
    \vspace{-5pt}
    \caption{Comparison of active power on bus 61 (generated pseudo-measurement) among different methods in a 24-h time horizon, under 95\% confidence level.}
    \label{fig:pseudo-mea predict}
    \vspace{-10pt}
\end{figure}

Fig.~\ref{fig:pseudo-mea predict} illustrates the active power estimation of the WTG at bus 61 with the associated uncertainty in a 24-h time horizon. 
Compared with two recent DNN-based methods, the proposed physics-bounded method accurately tracks the overall trend of the actual power output and yields more reliable estimates, especially during abrupt wind power variations. 
Moreover, it produces substantially tighter prediction intervals, indicating higher estimation accuracy. 
By contrast, the ANN-based method, which lacks physical constraints, may generate infeasible power ranges below 0, violating physical limits.

\begin{table}[!t]
\footnotesize                   
\setlength{\tabcolsep}{3pt}     
\captionsetup{
    font=small,
    justification=centering
}
\caption{\textsc{PDIN-inferred GMM Parameters for Three Buses in IEEE 123-Bus System
}}
\centering
\label{tab:gmm_123}
\begin{tabular*}{\linewidth}{@{\extracolsep{\fill}}|c|c|c|c|}
\hline
Location & $\boldsymbol{\pi}$ & $\boldsymbol{\mu}$ & $\boldsymbol{\sigma}$ \\ \hline
Bus 14 &
[0.2133, 0.7867] &
[0.1131, 0.1230] &
[0.0245, 0.0167] \\ \hline
Bus 61 &
\makecell{[0.0278, 0.9619,\\0.0102]} &
\makecell{[0.4568, 0.3071,\\0.4367]} &
\makecell{[0.0508, 0.0465,\\0.0913]} \\ \hline
Bus 151 &
\makecell{[0.0563, 0.9317,\\0.0120]} &
\makecell{[0.2698, 0.1984,\\0.3516]} &
\makecell{[0.0439, 0.0344,\\0.0755]} \\ \hline
\end{tabular*}
\end{table}
\begin{figure}[!t]
\centering
\includegraphics[width=0.37\textwidth]{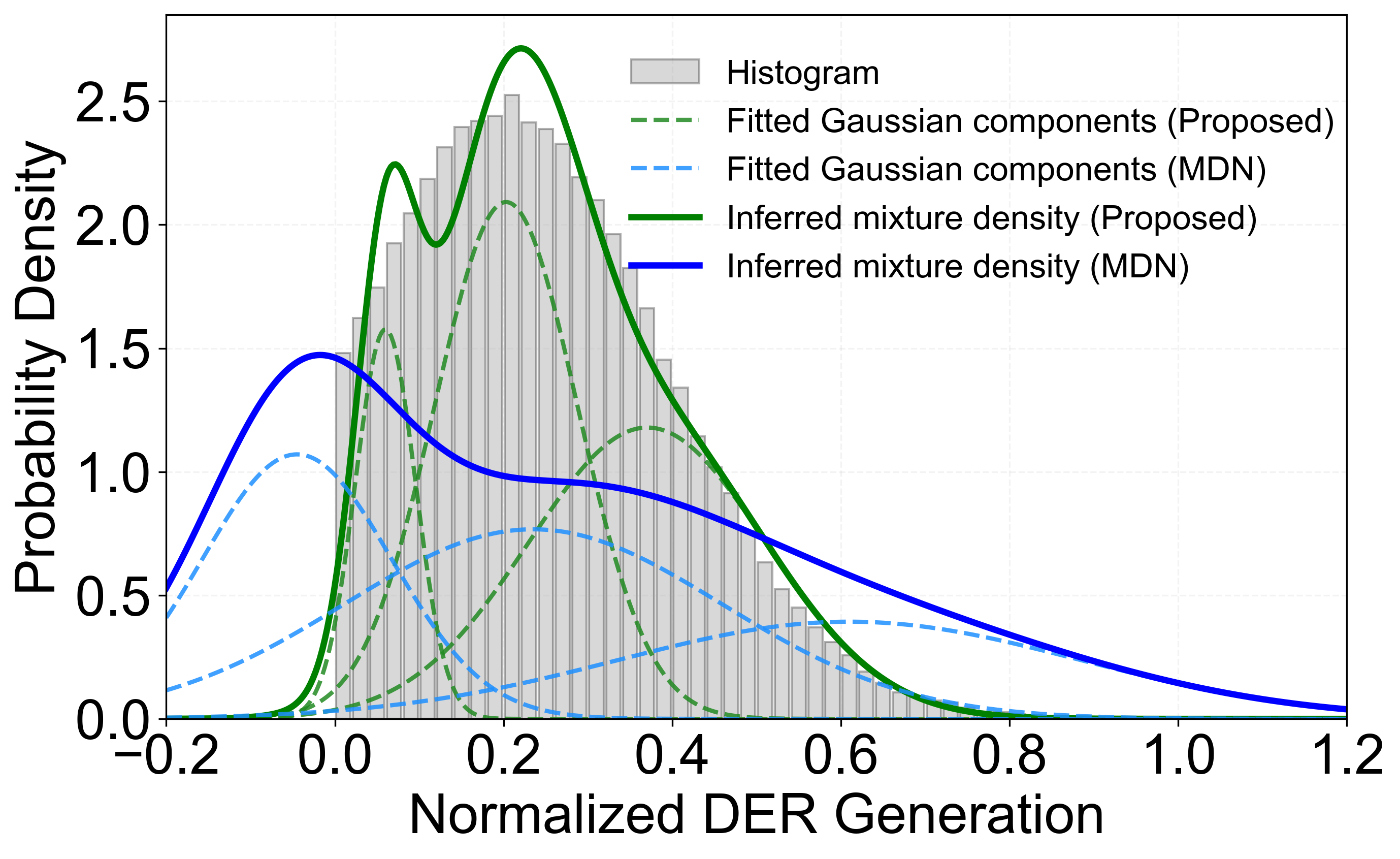}
\caption{
Illustrative comparison of the proposed dual-branch PDIN and single-branch MDN in the histogram of the WTG generation and inferred GMM approximations at bus 151 of the IEEE 123-bus system.
}
\label{fig:gmm_violin}
\end{figure}
Table \ref{tab:pdin} compares the MAEs and RMSEs of point estimates by different pseudo-measurement generation methods under non-Gaussian uncertainty. The proposed PDIN-based approach consistently outperforms both ANN and DBN baselines across all error measures.
Specifically, the MAEs of the ANN- and DBN-based methods are 0.0514 and 0.0278, respectively, whereas the proposed PDIN reduces the MAE to 0.0119, corresponding to a reduction of 76.9\% and 57.2\% compared with ANN and DBN, respectively.
Similar performance improvements are observed in terms of RMSE. The proposed PDIN achieves an RMSE of 0.0232, yielding error reductions of 66.4\% and 36.6\% relative to ANN and DBN, respectively.
These results indicate that the proposed PDIN is more effective in capturing the underlying non-Gaussian characteristics. 

The GMM parameters inferred by PDIN in the IEEE 123-bus system are summarized in Table~\ref{tab:gmm_123}.
Furthermore, based on these parameters, Fig.~\ref{fig:gmm_violin} compares the uncertainty distributions inferred by the proposed dual-branch PDIN and single-branch MDN. It is shown that the MDN-inferred distribution exhibits substantial density leakage into the negative region, which is physically infeasible for WTG generation. In comparison, the proposed PDIN constrains the inferred density within the feasible generation range and provides a closer approximation to the multimodal empirical distribution. This improvement is attributed to the proposed dual-branch architecture, which enhances the representation of complex non-Gaussian characteristics, and the physics-constrained loss functions, which explicitly enforce the physical generation boundaries during probabilistic density inference. Therefore, the proposed PDIN achieves improved non-Gaussian uncertainty representation while preserving the physical consistency of the inferred distribution.

\begin{figure}[!t]
    \centering
        \includegraphics[width=0.48\textwidth]{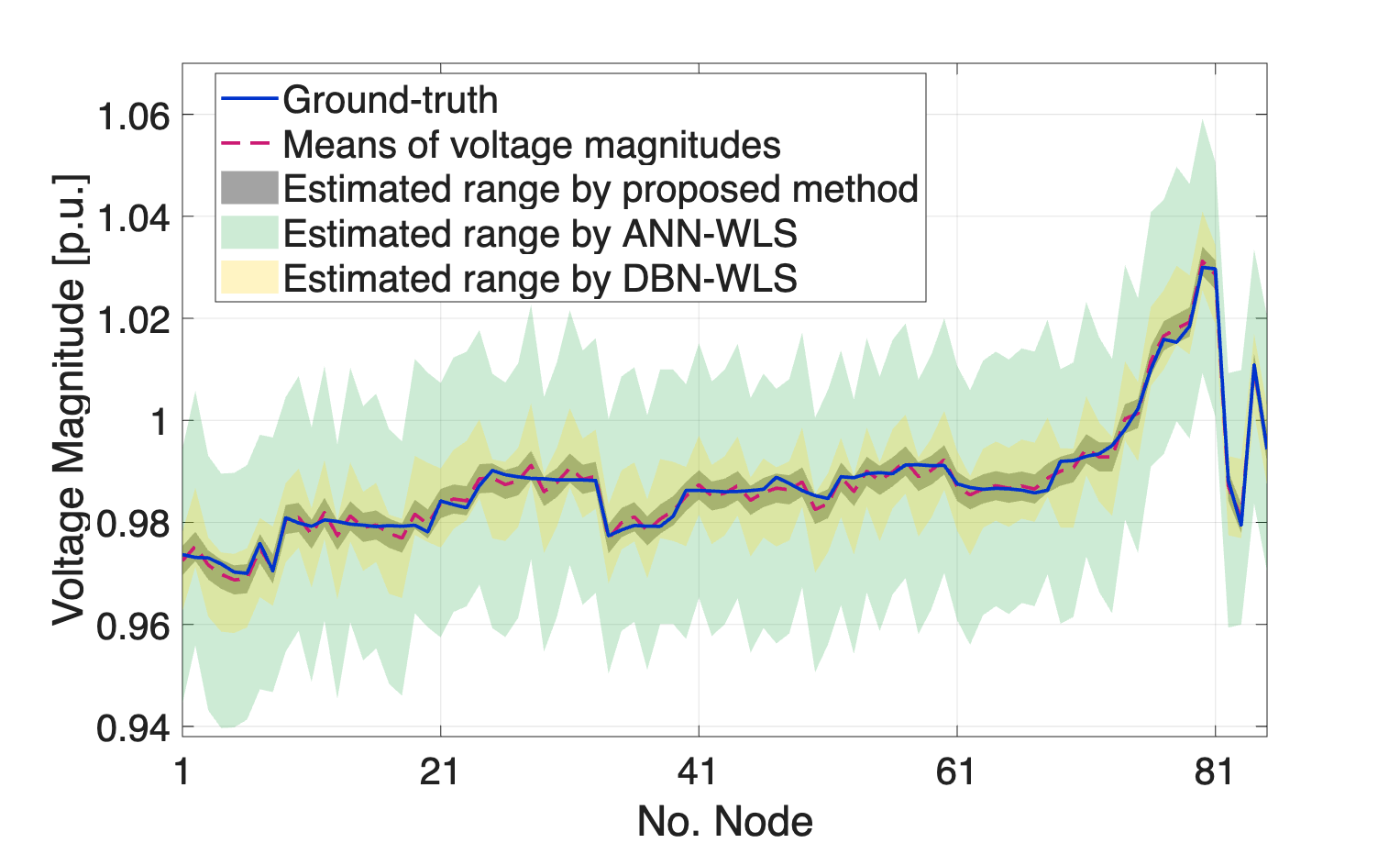}
    \vspace{-5pt}
    \caption{Comparison of voltage magnitude ranges estimated by different methods at noon. All the A-phase nodes are plotted in the 123-bus test system.}
    \label{fig:123voltageDist}
\end{figure}

\begin{table} 
\captionsetup{font=small,justification=centering}
\caption{\textsc{Comparison of Voltage Estimation Errors of Different DSSE Methods in 13-bus System}}
\label{tab:table1} 
\centering 
\small 
\begin{tabular}{|c|c|c|c|} 
\hline 
\multirow{2}{*}{{\makecell{Statistics\\Metrics}}}  
& \multirow{2}{*}{Method}  
& \multicolumn{2}{c|}{Voltage} \\ \cline{3-4}  
 &  & Magn. [p.u.] & Angle [rad] \\  
\hline 
\multirow{4}{*}{MAE}  
& WLS w/DGA  
& 2.01e-2 & 1.13e-1 \\  
& ANN Inference \cite{azimian2022state}  
&   8.15e-3 & 7.24e-2 \\  
& ANN-WLS \cite{manitsas2012distribution}  
& 7.27e-3 & 6.67e-2 \\  
& DBN-WLS \cite{huang2019probabilistic}  
& 5.81e-3 & 5.93e-2 \\  
& \textbf{Proposed}  
& \textbf{1.02e-3} & \textbf{2.65e-2} \\  
\hline 
\multirow{4}{*}{RMSE}   
& WLS w/DGA  
& 4.71e-2 & 4.20e-1 \\  
& ANN Inference \cite{azimian2022state}   
& 1.36e-2  & 1.14e-1 \\  
& ANN-WLS \cite{manitsas2012distribution}  
& 1.20e-2 & 1.03e-1 \\  
& DBN-WLS \cite{huang2019probabilistic}  
& 9.98e-3 & 8.78e-2 \\  
& \textbf{Proposed}  
& \textbf{1.67e-3} & \textbf{4.15e-2} \\  
\hline 
\multirow{4}{*}{{\makecell{Maximum\\Error}}}   
& WLS w/DGA  
& 8.62e-2 & 5.01e-1 \\ 
& ANN Inference \cite{azimian2022state}   
& 1.61e-2  &1.34e-1  \\  
& ANN-WLS \cite{manitsas2012distribution}  
& 1.43e-2 & 1.21e-1 \\  
& DBN-WLS \cite{huang2019probabilistic}  
& 9.73e-3 & 9.01e-2 \\  
& \textbf{Proposed}  
& \textbf{2.51e-3} & \textbf{4.77e-2} \\ 
\hline 
\end{tabular} 
\vspace{-5pt} 
\end{table}

\subsection{Comparison with Other DSSE Algorithms}

To assess the accuracy of voltage estimation, the proposed algorithm is compared with the WLS w/DGA, ANN-based direct state Inference, ANN-WLS, and DBN-WLS methods under non-Gaussian uncertainties.

Table~\ref{tab:table1} compares the voltage phasors estimated by different methods on the 13-bus system in terms of MAEs, RMSEs, and maximum errors. It is shown that the proposed method achieves the lowest error across all metrics, indicating consistently improved accuracy. For voltage magnitude, the proposed method reduces MAE to $1.02\times10^{-3}$ p.u., compared with $5.81\times10^{-3}$ p.u. for DBN-WLS, i.e., a 82.4\% reduction. For voltage angle, the MAE of the proposed method improves over DBN-WLS by 55.3\%, and over WLS w/DGA by 76.5\%. The WLS w/DGA method yields the highest errors because it simply neglects non-Gaussian uncertainties. Compared with the two-stage methods (i.e., ANN-WLS, DBN-WLS, and ours), ANN-based direct state inference without a dedicated uncertainty quantification stage exhibits higher voltage estimation errors.

\begin{figure}[!t]
    \centering
    \subfigure[]{
        \includegraphics[width=0.40\textwidth]{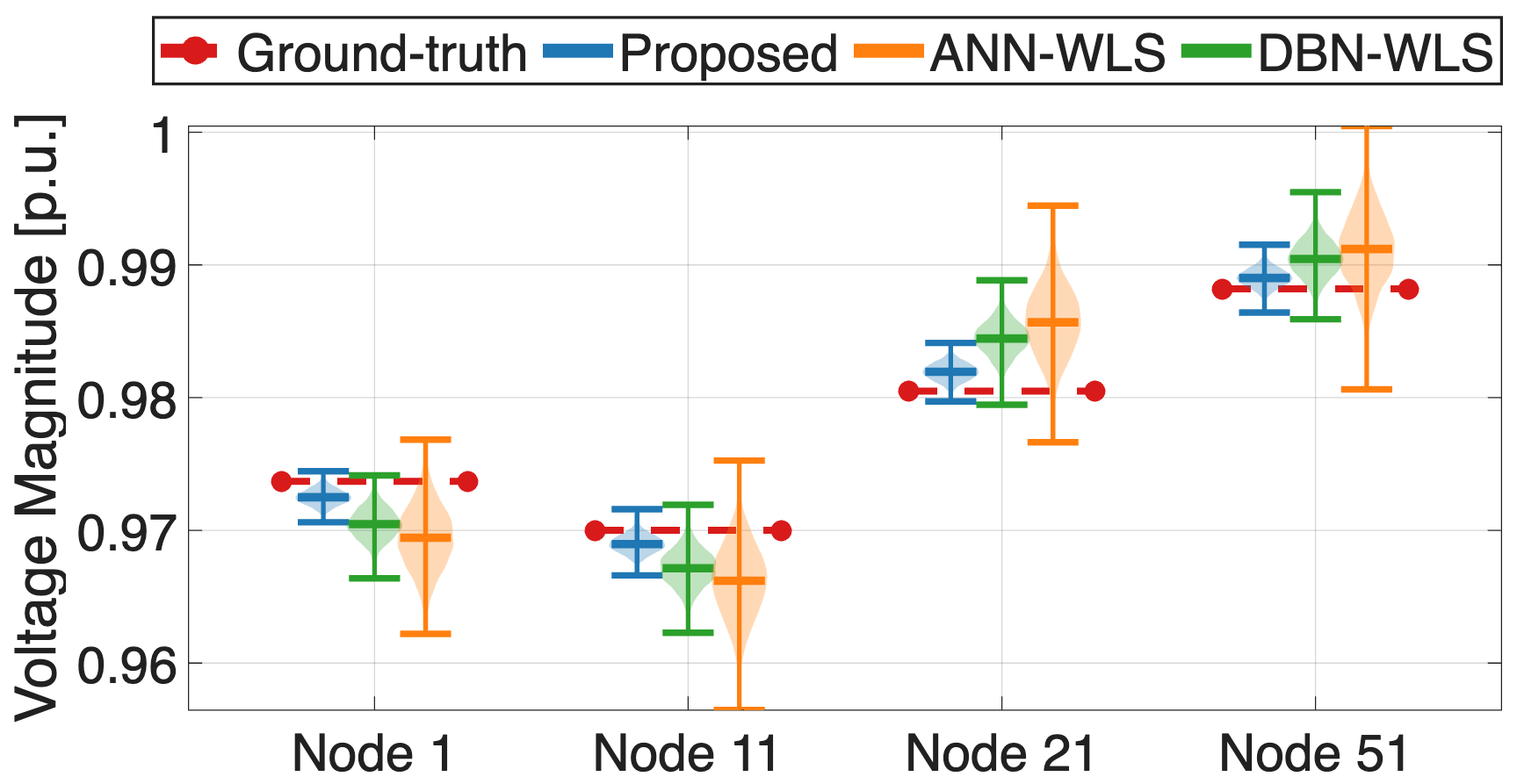}
    }\vspace{-5pt}
    \subfigure[]{
        \includegraphics[width=0.40\textwidth]{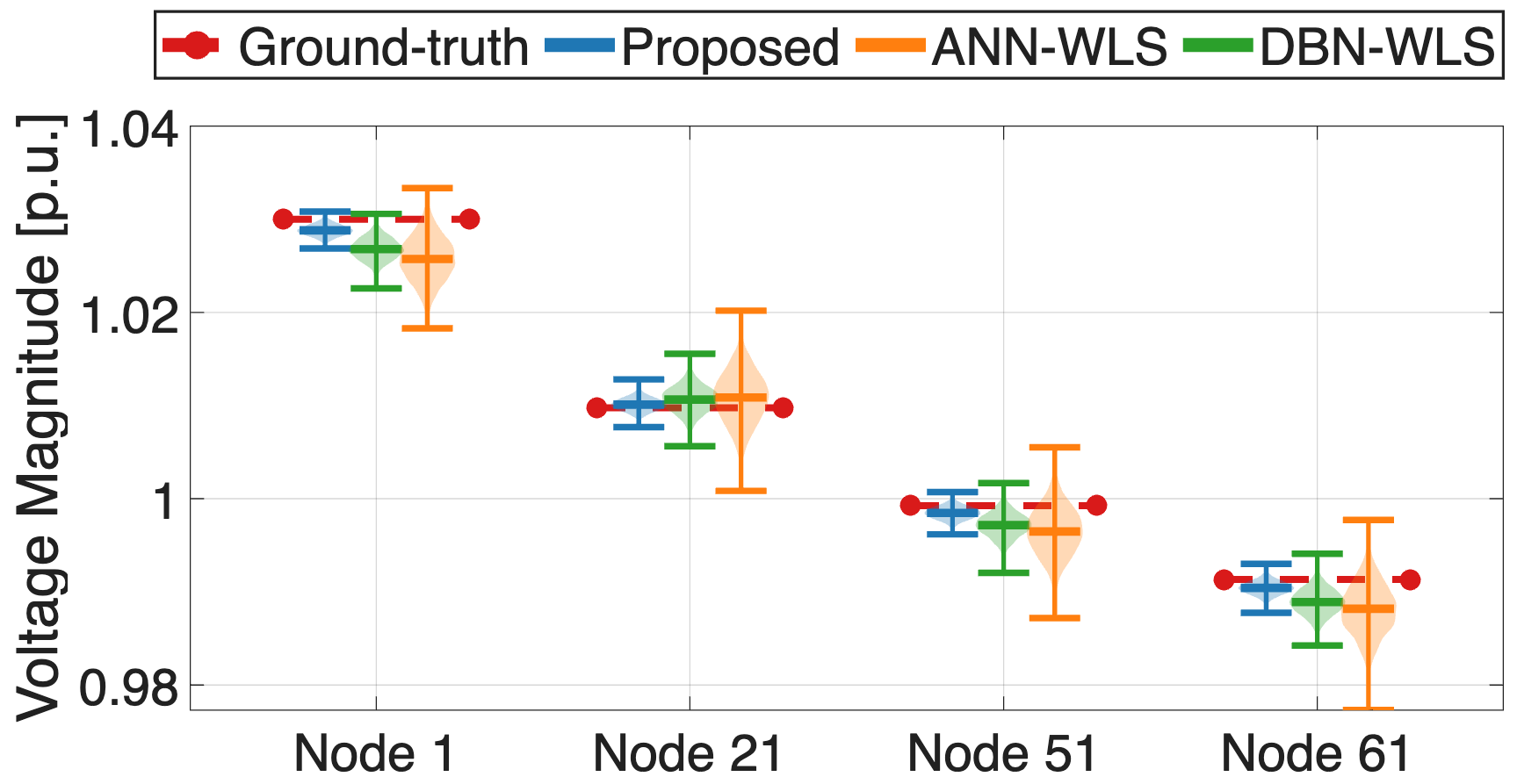}
    }\vspace{-10pt}
    \subfigure[]{
        \includegraphics[width=0.41\textwidth]{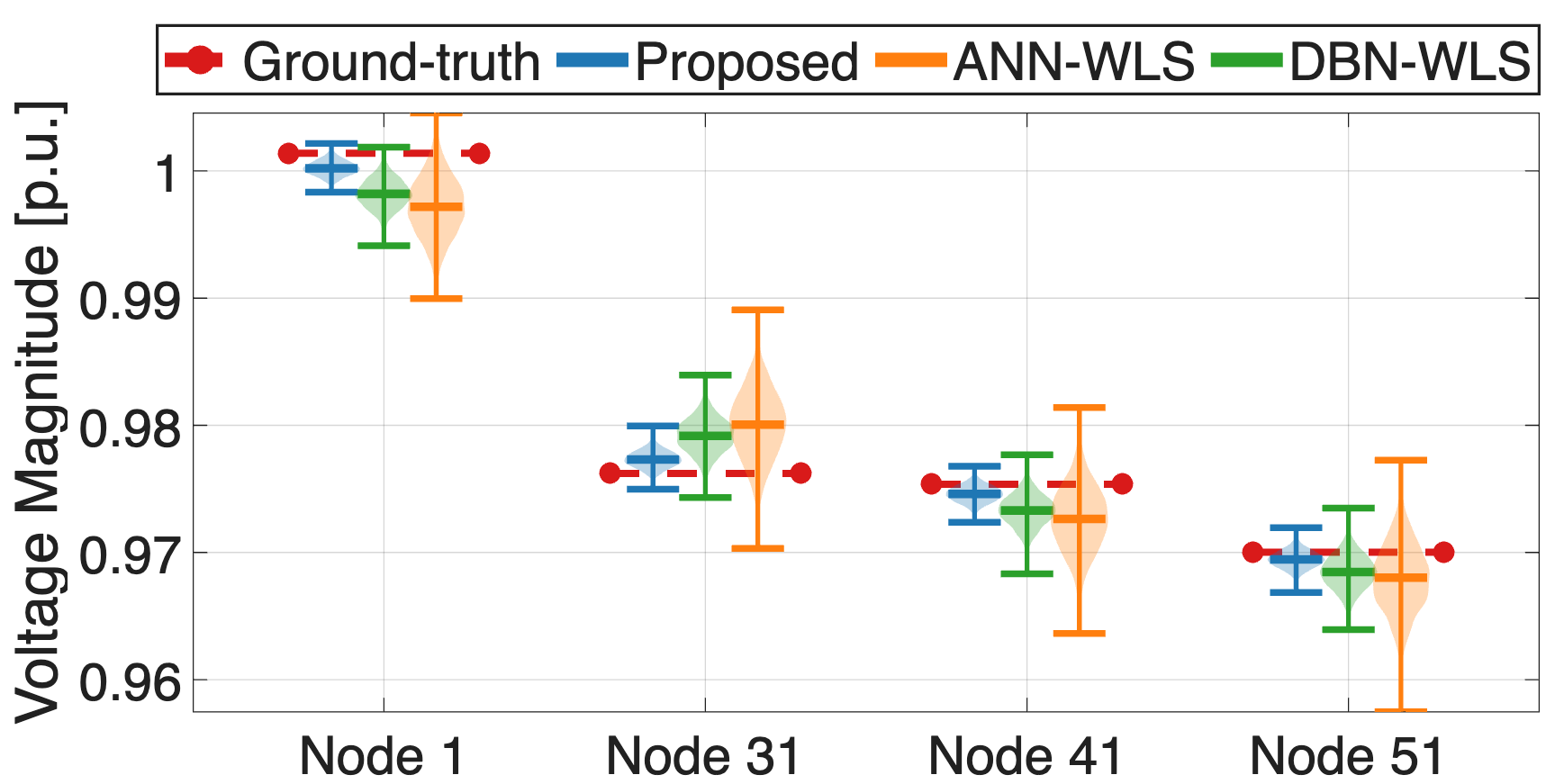}
    }
    \vspace{-10pt}
    \caption{Probability distributions of three-phase voltage magnitudes estimated by different methods in the 123-bus test system: (a) Phase A, (b) Phase B, (c) Phase C.}
    \label{fig:123voltage}
    \vspace{-14pt}
\end{figure}

\begin{figure}[!t]  
 \subfigure[]{
 \includegraphics[width=0.24\textwidth]{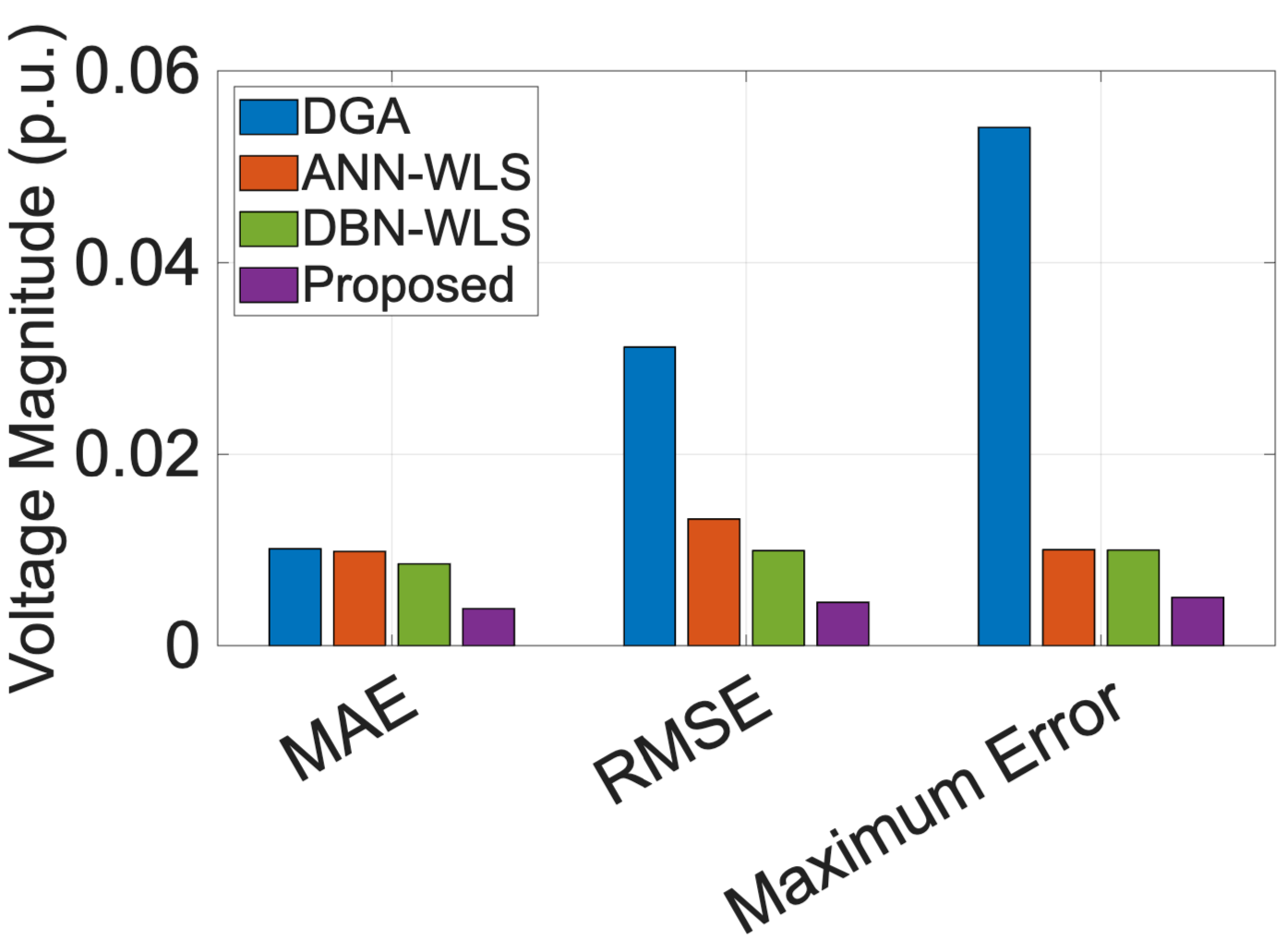}}\subfigure[]{
 \includegraphics[width=0.24\textwidth]{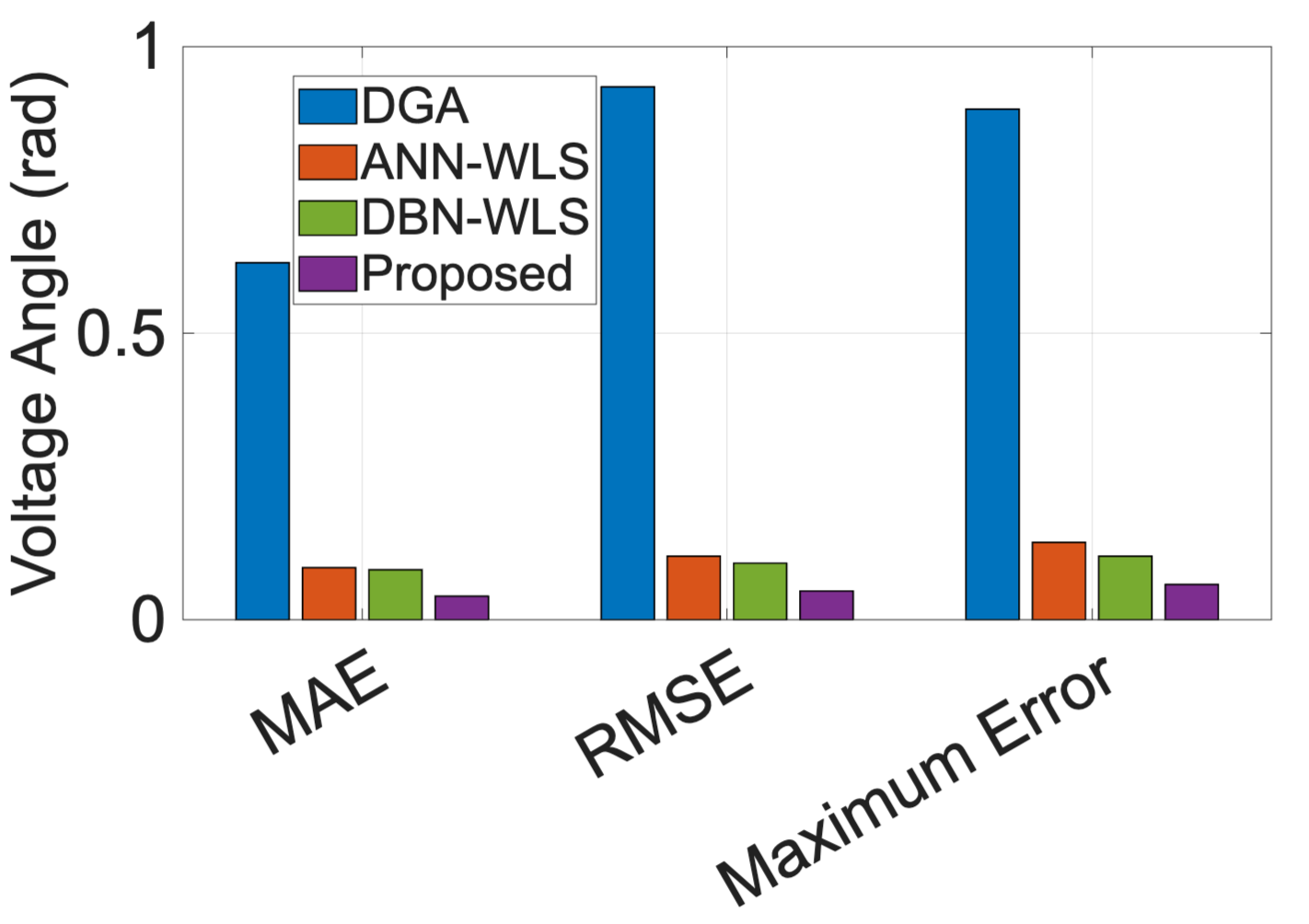}}
  \vspace{-5pt}
 \caption{MAEs, RMSEs, and maximum errors of different methods in the 123-bus system: (a) voltage magnitude, (b) voltage phase angle.}
 \label{fig:123result}
  \vspace{-10pt}
\end{figure}

The estimated distributions of A-phase voltage magnitudes obtained by different methods are shown for all the nodes in Fig.~\ref{fig:123voltageDist}. The proposed method produces significantly tighter uncertainty ranges that consistently cover the ground-truth values, indicating more accurate and less conservative uncertainty quantification. In contrast, DBN-WLS and ANN-WLS exhibit noticeably wider confidence intervals, particularly at downstream nodes with larger voltage variations, reflecting their limited ability to characterize non-Gaussian uncertainties.
Fig.~\ref{fig:123voltage} illustrates the three-phase voltage magnitude distributions at multiple buses in the 123-bus system. 
The proposed method yields estimates that are tightly concentrated around the ground truth, whereas the ANN-WLS and DBN-WLS show wider spreads and larger deviations.
It is shown that the voltage magnitude distributions of the proposed method remain consistently narrower and more concentrated than those of the benchmark methods across all nodes, demonstrating improved estimation stability.
The error results for the 123-bus test system in Fig. \ref{fig:123result} demonstrate that the proposed method consistently outperforms the other approaches with substantial improvements in estimation accuracy. For voltage magnitude, the MAE is reduced from $1.01\times10^{-2}$ for WLS w/DGA, $9.82\times10^{-3}$ for ANN-WLS, and $8.52\times10^{-3}$ for DBN-WLS to $3.85\times10^{-3}$, corresponding to an error reduction of roughly 55\%–65\% compared with the existing methods. A similar trend is observed for the RMSE metric, where the estimation error decreases from  $3.12\times10^{-2}$ for the WLS w/DGA to  $4.51\times10^{-3}$ with the proposed approach. Moreover, the maximum voltage magnitude error is reduced to $5.03\times10^{-3}$.
The MAE of voltage angles in the proposed method is approximately 55\% lower than those of ANN-WLS and DBN-WLS, and more than 90\% lower than that of WLS w/DGA. Consistent improvements are also observed in terms of RMSE and maximum error. These results confirm that the proposed estimator effectively suppresses large deviations induced by unknown non-Gaussian uncertainties.

\begin{table*}[!t] 
\centering 
\caption{\textsc{Comparison of Probabilistic Voltage Estimation Performance Under 95\% Confidence Level}}
\label{tab:probabilistic_results} 
\small 
\begin{tabular}{|c|c|c|c|c|c|c|c|} 
\hline 
\multirow{2}{*}{System} & 
\multirow{2}{*}{Method} & 
\multicolumn{2}{c|}{CRPS $\downarrow$} & 
\multicolumn{2}{c|}{PICP $\uparrow$} & 
\multicolumn{2}{c|}{MPIW $\downarrow$} \\  
\cline{3-8}
& & Magn. [p.u.] & Angle [rad] 
& Magn. [\%]  & Angle [\%] 
& Magn. [p.u.] & Angle [rad] \\ 
\hline 
\multirow{3}{*}{13-bus} 
& ANN-WLS 
& 5.21e-3 & 4.91e-2 & 91.8 & 91.2 & 2.84e-2 & 2.73e-1 \\ 
& DBN-WLS 
& 4.06e-3 & 4.27e-2 & 93.1 & 92.7 & 2.31e-2 & 2.41e-1 \\ 
& \textbf{Proposed} 
& \textbf{7.60e-4} & \textbf{1.92e-2} 
& \textbf{95.2} & \textbf{94.7} 
& \textbf{4.80e-3} & \textbf{1.12e-1} \\ 
\hline 
\multirow{3}{*}{123-bus} 
& ANN-WLS 
& 7.84e-3 & 6.36e-2 & 90.9 & 90.4 & 3.72e-2 & 3.16e-1 \\ 
& DBN-WLS 
& 6.12e-3 & 5.48e-2 & 92.5 & 91.9 & 3.08e-2 & 2.78e-1 \\ 
& \textbf{Proposed} 
& \textbf{2.71e-3} & \textbf{3.08e-2} 
& \textbf{94.5} & \textbf{94.1} 
& \textbf{1.42e-2} & \textbf{1.68e-1} \\ 
\hline 
\multicolumn{8}{l}{\footnotesize *Lower CRPS and MPIW values indicate better performance, while a higher PICP indicates better prediction interval coverage. } 
\end{tabular} 
\vspace{-5pt} 
\end{table*}
Table~\ref{tab:probabilistic_results} further evaluates and compares the probabilistic
voltage estimation performance in terms of CRPS, PICP, and MPIW, under the nominal 95\% confidence level. An accurate probabilistic estimator is
expected to achieve a PICP close to the nominal coverage while maintaining
a narrow prediction interval. It is shown that the proposed method consistently achieves the lowest CRPS and MPIW for both voltage magnitude and phase angle in the two systems, while maintaining PICP close to the nominal coverage level. These results demonstrate improved distributional accuracy and sharper uncertainty intervals without sacrificing the required coverage. For the 13-bus test system, the proposed method
achieves PICP values of 95.2\% and 94.7\% for voltage magnitude and angle,
respectively, compared with 91.8\% and 91.2\% for ANN-WLS and 93.1\% and
92.7\% for DBN-WLS. Similar performance is observed in the 123-bus system,
where the proposed method maintains PICP values of 94.5\% and 94.1\%,
which remain closer to the nominal 95\% level than those of ANN-WLS and
DBN-WLS. 
% Although a slight
% performance degradation is observed in the 123-bus system due to its
% increased system dimension and uncertainty propagation complexity, the
% proposed method maintains reliable probabilistic estimation performance.
Overall, these results demonstrate that the proposed method significantly improves not
only point-estimation accuracy but also the calibration and sharpness of
the probabilistic voltage estimates.

\subsection{Computational Time for Online Voltage Estimation}
Table \ref{tab:time} lists the online computational time in seconds (s)
for the proposed learning-enabled voltage estimation with unknown uncertainty modeling and the nonlinear estimators.

In the 13-bus test system, the proposed method achieves an average solution time of 0.04~s, which is approximately more than 7.8 times faster than the ANN-WLS and DBN-WLS approaches. Similar performance gains are observed in the 123-bus system, where the proposed method completes the estimation within 0.57~s, compared with over 8.2~s for both ANN-WLS and DBN-WLS. This corresponds to a speed-up factor of more than one order of magnitude. It should be noted that the proposed algorithm directly estimates the probability distribution of nodal voltages via a single Monte Carlo trial and non-iterative solving.
In contrast, the number of iterations required to solve the recent nonlinear estimators significantly increases, reaching an average of 72 times. It is evident that the CPU time of the proposed BBL-enabled estimator on its online implementation is almost unaffected by the distribution system size and non-Gaussian input statistics.

\begin{table}[!t]
\centering
\captionsetup{font=small,justification=centering}
\caption{\textsc{Comparison of Average CPU Time}}
\label{tab:time}
\centering
\small
\begin{tabular}{|c|c|c|c|c|}
\hline
\multirow{2}{*}{\# Method} & \multicolumn{2}{c|}{13-Bus System} & \multicolumn{2}{c|}{123-Bus System} \\ \cline{2-5} 
 & \makecell{Iter. \\ Counts} & \makecell{CPU \\ Time [s]} & \makecell{Iter. \\ Counts} & \makecell{CPU \\ Time [s]} \\ \hline
ANN-WLS \cite{manitsas2012distribution} & 12.7 & 0.26 & 72.1 & 8.22 \\ \hline
DBN-WLS \cite{huang2019probabilistic}  & 12.1 & 0.27 & 71.3 & 8.23 \\ \hline
\textbf{Proposed} & \textbf{N/A} & \textbf{0.04} & \textbf{N/A} & \textbf{0.57} \\ \hline
\multicolumn{5}{l}{\footnotesize *Only two-stage hybrid DSSE methods that can handle unknown } \\
\multicolumn{5}{l}{\footnotesize non-Gaussian uncertainty are compared here.}
\end{tabular}
\end{table}
\subsection{Evaluating Proposed Physics-Guided Design in Stage Two}
This subsection evaluates the contribution of the proposed BBL in Stage Two through an ablation study, in which only BBL is replaced with a deterministic ResNN. The ResNN uses the same input, hidden-layer architecture, training data, and data-fitting term as BBL, but removes the Bayesian weight distributions and KL-divergence regularization. Therefore, the comparison isolates the contribution of Bayesian learning to voltage estimation accuracy and uncertainty quantification while preserving the same residual bias-correction architecture and physics-guided design.

\begin{figure}[!t]
    \centering
        \includegraphics[width=0.43\textwidth]{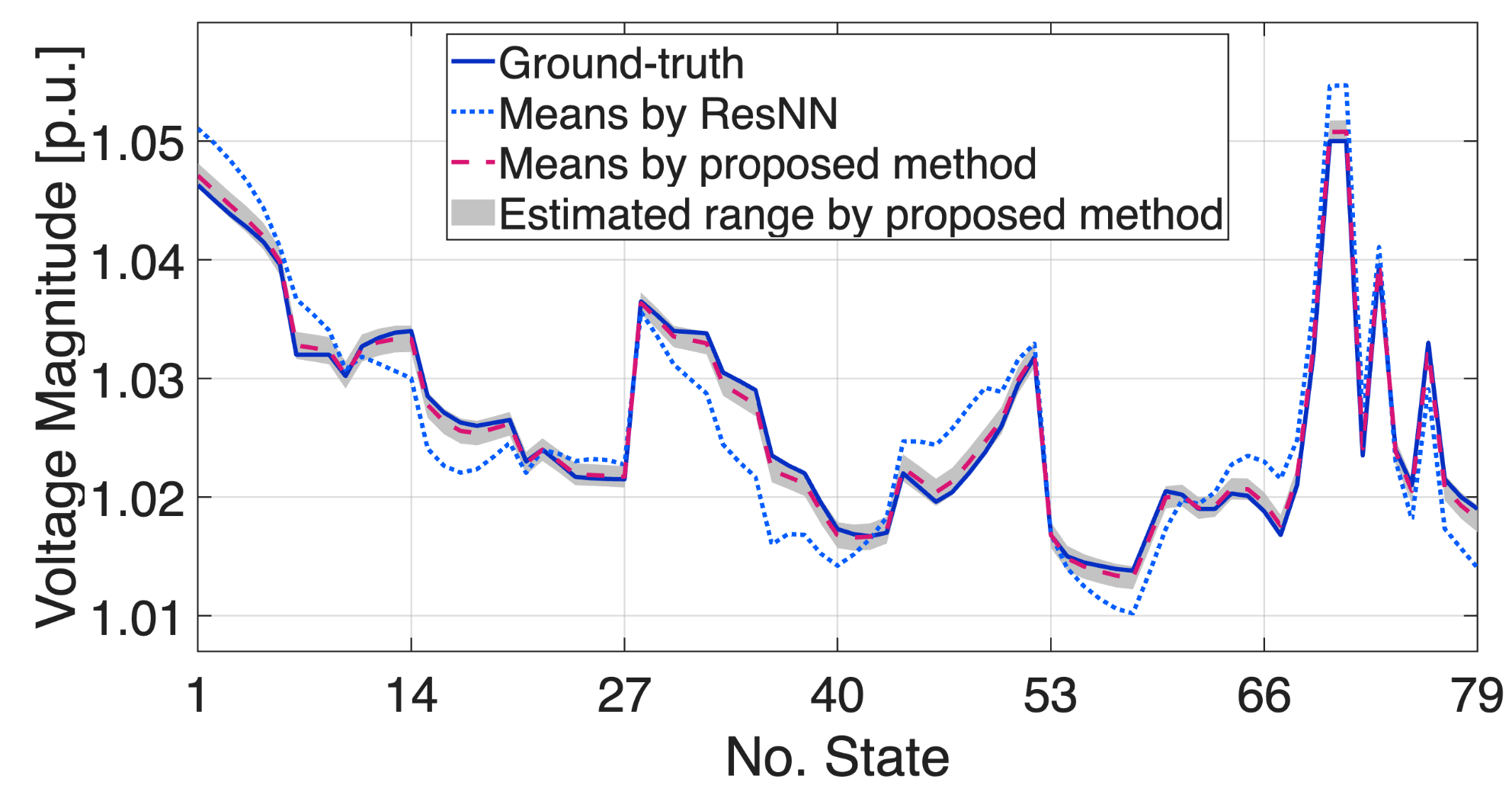}
    \vspace{-5pt}
    \caption{Comparison with ResNN-based state estimation in voltage magnitude. All the B-phase voltages in the 123-bus test system are plotted.}
    \label{fig:rangemag}
\end{figure}
% \begin{figure}[!t]
%     \centering
%         \includegraphics[width=0.48\textwidth]{image/bangle.png}
%     \vspace{-5pt}
%     \caption{\textcolor{blue}{Comparison with ResNN-based state estimation in phase angle ranges. All the B-phase voltages are plotted in the 123-bus test system.}}
%     \label{fig:rangeangle}
% \end{figure}
\begin{table}[!t]
\captionsetup{font=small,justification=centering}
\caption{\textsc{Comparison of BBL and ResNN in IEEE 13- and 123-Bus Systems}}
\label{tab:bbl_drnn}
\centering
\small
\begin{tabular}{|c|c|c|c|}
\hline
\multirow{2}{*}{{\makecell{Statistics\\Metrics}}}
& \multirow{2}{*}{Method}
& \multicolumn{2}{c|}{Voltage} \\ \cline{3-4}
& & Magn. [p.u.] & Angle [rad] \\
\hline
\multicolumn{4}{|c|}{13-Bus Distribution System} \\
\hline
\multirow{2}{*}{MAE}
& PDIN+ResNN
& 2.36e-3
&3.71e-2 \\
& \textbf{Proposed}
& \textbf{1.02e-3}
& $\textbf{2.65e-2}$ \\
\hline
\multirow{2}{*}{RMSE}
& PDIN+ResNN
& 3.84e-3
& 5.52e-2 \\
& \textbf{Proposed}
& \textbf{1.67e-3}
& \textbf{4.15e-2} \\
\hline
\multirow{2}{*}{{\makecell{Maximum\\Error}}}
& PDIN+ResNN
& 5.92e-3
& 6.83e-2 \\
& \textbf{Proposed}
& \textbf{2.51e-3}
& \textbf{4.77e-2} \\
\hline
\multicolumn{4}{|c|}{123-Bus Distribution System} \\
\hline
\multirow{2}{*}{MAE}
& PDIN+ResNN
& 4.86e-3
& 5.21e-2 \\
& \textbf{Proposed}
& \textbf{3.80e-3}
& \textbf{4.30e-2} \\
\hline
\multirow{2}{*}{RMSE}
& PDIN+ResNN
& 5.94e-3
& 6.43e-2 \\
& \textbf{Proposed}
& \textbf{4.50e-3}
& \textbf{5.20e-2} \\
\hline
\multirow{2}{*}{{\makecell{Maximum\\Error}}}
& PDIN+ResNN
& 7.18e-3
& 7.86e-2 \\
& \textbf{Proposed}
& \textbf{5.00e-3}
& \textbf{6.50e-2} \\
\hline
\end{tabular}
\end{table}
Fig. \ref{fig:rangemag} shows that the proposed method provides substantially narrower estimation ranges while closely tracking the ground truth for voltage magnitude across the phase B nodes. In contrast, ResNN provides only deterministic point estimates, which exhibit larger deviations from the ground truth and do not provide uncertainty quantification. These results demonstrate that the proposed method achieves improved voltage estimation accuracy while providing uncertainty-aware state estimates.

As shown in Table \ref{tab:bbl_drnn}, compared with the ResNN counterpart, the proposed BBL-enabled estimator reduces the voltage 
magnitude and phase angle MAEs by 56.8\% and 28.6\%, respectively, 
in the 13-bus system. This comparison demonstrates that 
the performance improvement is not solely attributed to the residual-learning 
architecture, but also benefits from Bayesian learning.
For the 123-bus system, the proposed method consistently outperforms 
ResNN across all three error metrics. In particular, the voltage magnitude 
and phase angle MAEs are reduced by 21.8\% and 17.5\%, respectively, 
demonstrating that the benefit of Bayesian bias learning is maintained 
in the larger distribution system.

Overall, the proposed BBL-enabled estimator consistently achieves lower 
MAE, RMSE, and maximum errors than its deterministic counterpart in both 
test systems. Since the two methods share the same analytical estimator, 
network architecture, inputs, and training data, these results demonstrate 
the effectiveness of Bayesian learning in improving voltage state estimation 
beyond deterministic residual bias correction.

\subsection{Sensitivity Analysis}
\begin{figure}[!t]
    \centering
        \subfigure[Sensitivity on M]{ \includegraphics[width=0.48\textwidth]{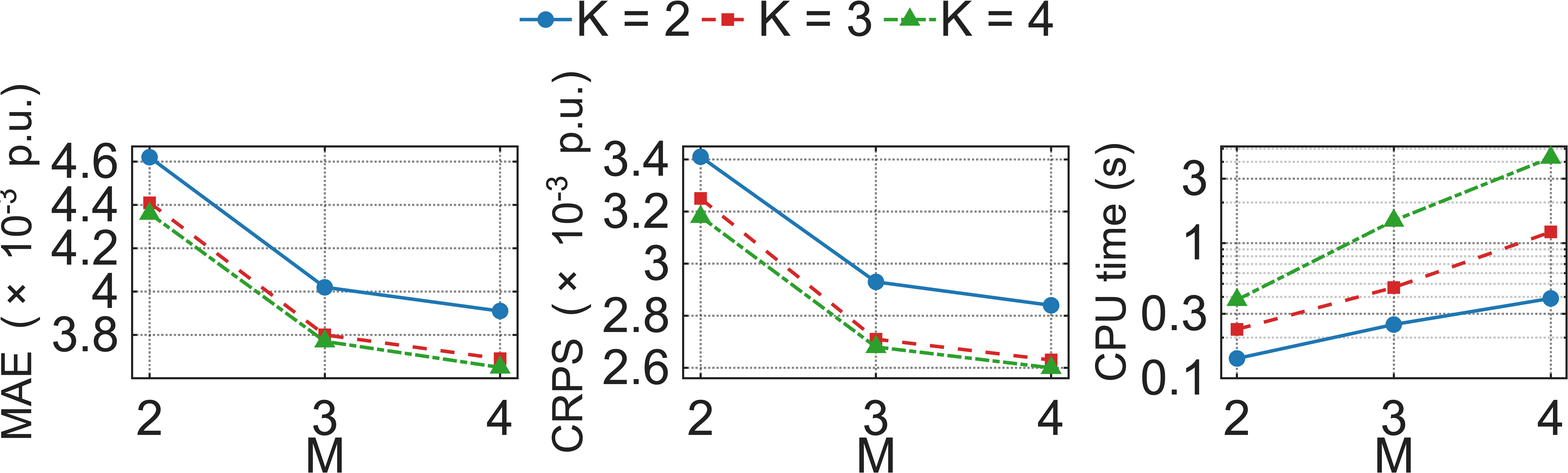}}
         \subfigure[Sensitivity on K]{
        \includegraphics[width=0.48\textwidth]{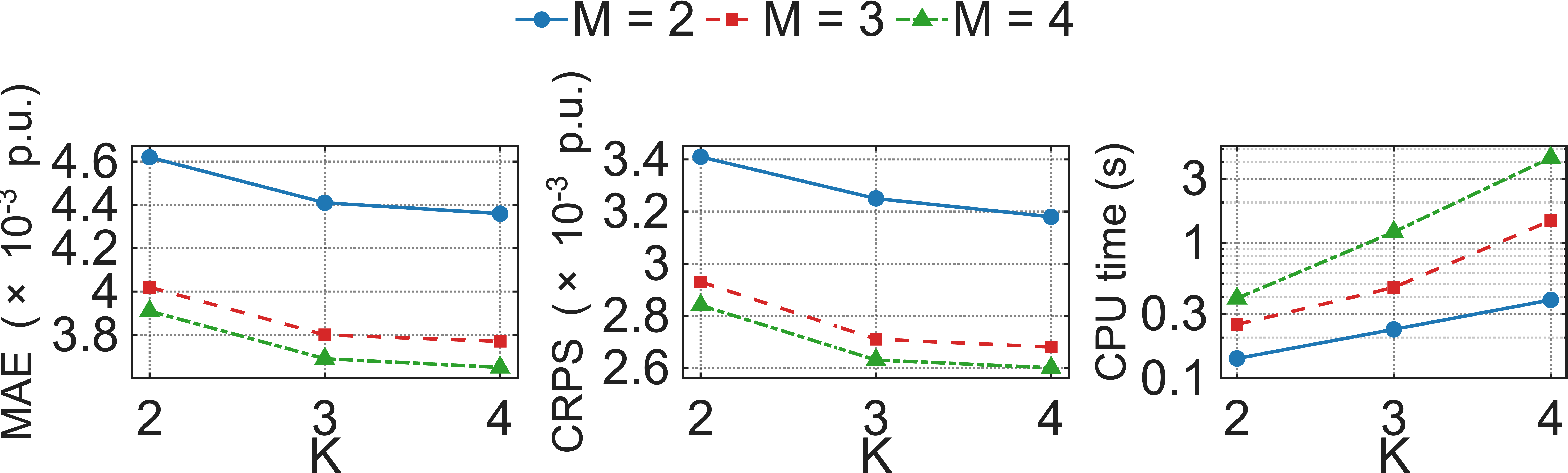}}
    \vspace{-5pt}
    \caption{Voltage estimation errors and CPU time for different M and K in the IEEE 123-bus system.}
    \label{fig:mksensitity}
\end{figure}

The impact of the number of DER uncertainty sources $M$ and the number of GMM
components $K$ on the computational cost and estimation errors of the proposed
method is evaluated.
Fig. \ref{fig:mksensitity} shows the MAE, CRPS, and CPU time under different combinations of $M$ and $K$ in the IEEE 123-bus distribution system. 
% We select the value of $K$ ranging from 2 to 4, as \cite{singh2010disGMM} shows that an excessive number of Gaussian components may be redundant and similar components can be merged through mixture reduction.

In Fig. \ref{fig:mksensitity}a), for a fixed $M$, increasing $K$ from 2 to 3 noticeably reduces both MAE
and CRPS, indicating improved estimation accuracy due to the finer representation of the non-Gaussian DER
uncertainty. 
Meanwhile, increasing the number of GMM components $K$ also increases the computational cost.
For example, when $M=3$, as $K$ increases from 2 to 3, the average CPU time increases from about 0.23 seconds to 0.47 seconds, while the voltage magnitude MAE decreases from $4.41\times10^{-3}$ p.u. to $3.80\times10^{-3}$ p.u. When further increasing $K$ from 3 to 4, the average CPU time increases from 0.47 seconds to 1.21 seconds, whereas the voltage magnitude MAE decreases only slightly from $3.80\times10^{-3}$ p.u. to $3.69\times10^{-3}$ p.u. The proper selection of $K$ can be pre-determined by clustering \cite{5298967GMM}; an excessive number of Gaussian components may be redundant, as similar components can be merged through mixture reduction \cite{singh2010disGMM}.

Although the number of DSSE subproblems grows
exponentially as $K^M$, \textcolor{blue}{the computation time of the proposed estimator remains low, compared with the existing methods. Each subproblem is solved by a linear analytical solution followed by a BBL forward pass for bias correction, so no iterative state updates are required. In addition, the subproblems are solved in parallel to reduce the elapsed time. As reported in
Fig.~\ref{fig:mksensitity}, the proposed method only
takes 1.21~s per estimate with $(M,K)=(3,4)$ in the IEEE 123-bus system, achieving near real-time monitoring.}

% \textcolor{blue}{To determine how many subproblems can be processed in an online
% application, an illustrative budget of 1~s per estimate is adopted,
% corresponding to one state estimation run per cycle in fast
% distribution voltage control \cite{xujian}. 
% As shown in Fig.~\ref{fig:mksensitity}, the configurations
% $(M,K)=(2,2)$, $(3,2)$, $(4,2)$, $(2,4)$, and $(3,3)$
% satisfy the 1-s computational budget, with a maximum of
% 27 subproblems at $(M,K)=(3,3)$.
% }

\textcolor{blue}{For systems with a largerer number of uncertain DERs, the scalability can be further improved within the proposed framework in three ways. First, the GMM representation can be limited to DERs with strong non-Gaussian uncertainty, and the remaining DERs are represented by a single Gaussian component. A single-component source does not increase the number of combinations, so the number of subproblems becomes $\prod_{i=1}^{M}K_i$, where $K_i$ is the number of Gaussian components of the $i$-th source. Second, $K_i$ can be reduced by merging similar components through mixture reduction \cite{singh2010disGMM}. Third, the combinations with negligible weights $w_c$ in \eqref{eq:x_final} can be discarded, and the remaining weights are renormalized.
% These strategies reduce the computation time at the cost of an approximation error.
}

\section{Conclusion}
This paper proposes a BBL-enabled probabilistic voltage estimation in unbalanced distribution systems under unknown uncertainties. The proposed PDIN-based method jointly models pseudo-measurements and their probabilistic distributions, enabling effective quantification of non-Gaussian uncertainties arising from DERs. By incorporating the resulting probabilistic distribution as pseudo-measurements into a physics-informed BBL-enabled linear estimator, the proposed two-stage framework achieves uncertainty-aware, scalable voltage estimation.
Case studies demonstrate that our method achieves superior accuracy and computational efficiency, compared with recent data-driven and model-based DSSE methods. The proposed probabilistic uncertainty modeling can further support reliable grid operation and control under unknown uncertainties in active distribution systems with high DER penetration. 
 Future work will extend the proposed uncertainty modeling
framework to correlated DER uncertainties by developing a joint 
PIDN-based probabilistic inference approach that outputs extra correlation parameters for geographically close distributed generation units. The resulting 
cross-covariance information can be incorporated into the non-diagonal locations of the 
measurement covariance matrix in \eqref{Covar}, allowing the proposed voltage estimator 
to account for correlated uncertainties. 

\appendix
Given a specified combination of sensors or pseudo-measurements, the linearized formulations for available power injection, power flow, and voltage magnitude measurements used in the proposed estimation method are presented below to obtain $\mathbf{H}_l$:

{1)  Power Injection Measurements} 

The power measurements at node $i$ hold a nonlinear relationship
\begin{equation}\label{powerinjection}
\mathbf s_i=\mathbf p_i+j\mathbf q_i=\mathbf v_i \sum_{l \in N(i)}[\mathbf Y_{ij} (\mathbf v_i-\mathbf v_j)]^*      
\end{equation}
where $\mathbf p_i$  and $\mathbf q_i$  denote the multi-phase real and reactive powers at bus $i$, and $N(i)$ is the set of all buses connected to bus $i$; $\mathbf{Y}_{ij} = \mathbf{G}_{ij} + j\mathbf{B}_{ij}$ represents the multi-phase admittance matrix between bus $i$ and bus $j$, where $\mathbf{G}_{ij}$ and $\mathbf{B}_{ij}$ are the conductance and susceptance matrices, respectively. The operator $[\cdot]^*$ denotes the complex conjugate.

 By approximating $\mathbf v_i$ with the substation voltage measurement $\mathbf V_0$ in \eqref{powerinjection}, the following linear measurement function is obtained: 
\begin{equation}\label{16}
 \mathbf h_{p,i}+j\mathbf h_{q,i}= \mathbf V_0\sum_{j \in \mathcal{N}(i)} [{\mathbf Y_{i j}}\left(\mathbf{v}_i-\mathbf{v}_j\right)]^*
\end{equation}
where $\mathbf h_{p,i}(\mathbf x)$ and $\mathbf h_{q,i}(\mathbf x)$ denote the measurement functions for power injection on bus $i$. Interested readers can refer to \cite{LinearPF7031975} for the error analysis of this linear approximation. 

The Jacobian elements of the power injections w.r.t. the voltage states at buses $i$ and $j$ are given by:

\begin{equation}
\frac{\partial \mathbf h_{p,i}}{\partial \mathbf v_{ir}}=\mathbf V_0\sum_{j \in N(i)}\mathbf G_{ij}, \quad
\frac{\partial \mathbf h_{p,i}}{\partial \mathbf v_{ix}}=-\mathbf V_0 \sum_{j \in N(i)}\mathbf B_{ij}
\end{equation}

\begin{equation}
\frac{\partial \mathbf h_{p,i}}{\partial \mathbf v_{jr}}=-\mathbf V_0 \mathbf G_{ij}, \quad
\frac{\partial \mathbf h_{p,i}}{\partial \mathbf v_{jx}}=\mathbf V_0 \mathbf B_{ij}
\end{equation}
\begin{equation}
\frac{\partial \mathbf h_{q,i}}{\partial \mathbf v_{ir}}=-\mathbf V_0 \sum_{j \in N(i)}\mathbf B_{ij}, \quad
\frac{\partial \mathbf h_{q,i}}{\partial \mathbf v_{ix}}=-\mathbf V_0 \sum_{j \in N(i)}\mathbf G_{ij}
\end{equation}
\begin{equation}
\frac{\partial \mathbf h_{q,i}}{\partial \mathbf v_{jr}}=\mathbf V_0 \mathbf B_{ij}, \quad
\frac{\partial \mathbf h_{q,i}}{\partial \mathbf v_{jx}}=\mathbf V_0 \mathbf G_{ij}
\end{equation}
where  
$\mathbf v_{ir}$ and $\mathbf v_{ix}$ denote the real and imaginary parts at node $i$, respectively; $\mathbf v_{jr}$ and $\mathbf v_{jx}$ denote the real and imaginary parts of the voltage at node $j$, respectively.

{2)  Power Flow Measurements}
Similarly, the linearized measurement functions for the power flow measured at branch $i$-$k$, denoted as $\mathbf h_{pl,ik}(\mathbf x)$ and $\mathbf h_{ql,ik}(\mathbf x)$, are approximated by replacing $\mathbf v_i$ with $\mathbf V_0$: 
\begin{equation}\label{18}
\mathbf h_{pl,ik}+j\mathbf h_{ql,ik}=\mathbf V_0 [\mathbf Y_{ik}(\mathbf v_i-\mathbf v_k)]^*
\end{equation}

{3) Voltage Magnitude Measurements}

The measurement function of $\phi$-phase voltage magnitudes at bus $k$ is approximated as a linear relationship between $|v_k^\phi|^2$ and voltage variables:
\begin{equation} \label{volatgemagnitude}
|v_k^\phi|^2=(v_{k,r}^\phi)^2+(v_{k,x}^\phi)^2
\approx V_{0,r}^\phi v_{k,r}^\phi+V_{0,x}^\phi v_{k,x}^\phi    
\end{equation}
where $v_{k,r}^\phi$  and $v_{k,x}^\phi$ represent the real and imaginary parts of the voltage state at bus $k$.

The measurement function for the voltage magnitude, denoted as $\mathbf h_{|v_k^\phi|^2}$, is developed by a linearized function:
\begin{equation} \label{20}
\mathbf h_{|v_k^\phi|^2}(\mathbf x)=V_{0,r}^\phi v_{k,r}^\phi+V_{0,x}^\phi v_{k,x}^\phi    
\end{equation}
where $V_{0,r}^\phi$ and $V_{0,x}^\phi$ denote the real and imaginary parts of the substation voltage measurement.
The non-zero Jacobian elements of the measurement function \eqref{20} are expressed as:
\begin{equation}
\frac{\partial \mathbf h_{|v_k^\phi|^2}}{\partial v_{k,r}^\phi}=V_{0,r}^\phi,\quad
\frac{\partial \mathbf h_{|v_k^\phi|^2}}{\partial v_{k,x}^\phi}=V_{0,x}^\phi
\end{equation}

\IEEEpubidadjcol

\bibliographystyle{IEEEtran}
\bibliography{IEEEabrv,Citation}

% Generated by IEEEtran.bst, version: 1.14 (2015/08/26)
\begin{thebibliography}{10}
\providecommand{\url}[1]{#1}
\csname url@samestyle\endcsname
\providecommand{\newblock}{\relax}
\providecommand{\bibinfo}[2]{#2}
\providecommand{\BIBentrySTDinterwordspacing}{\spaceskip=0pt\relax}
\providecommand{\BIBentryALTinterwordstretchfactor}{4}
\providecommand{\BIBentryALTinterwordspacing}{\spaceskip=\fontdimen2\font plus
\BIBentryALTinterwordstretchfactor\fontdimen3\font minus \fontdimen4\font\relax}
\providecommand{\BIBforeignlanguage}[2]{{%
\expandafter\ifx\csname l@#1\endcsname\relax
\typeout{** WARNING: IEEEtran.bst: No hyphenation pattern has been}%
\typeout{** loaded for the language `#1'. Using the pattern for}%
\typeout{** the default language instead.}%
\else
\language=\csname l@#1\endcsname
\fi
#2}}
\providecommand{\BIBdecl}{\relax}
\BIBdecl

\bibitem{pse7312455}
A.~Angioni, T.~Schlösser, F.~Ponci, and A.~Monti, ``Impact of pseudo-measurements from new power profiles on state estimation in low-voltage grids,'' \emph{IEEE Transactions on Instrumentation and Measurement}, vol.~65, no.~1, pp. 70--77, 2016.

\bibitem{singh2010disGMM}
R.~Singh, B.~Pal, and R.~Jabr, ``Distribution system state estimation through gaussian mixture model of the load as pseudo-measurement,'' \emph{IET Generation, Transmission \& Distribution}, vol.~4, no.~1, pp. 50--59, 2010.

\bibitem{zhangInterval}
Y.~Zhang{ }, J.~Wang, and Z.~Li, ``Interval state estimation with uncertainty of distributed generation and line parameters in unbalanced distribution systems,'' \emph{IEEE Trans. Power Systems}, vol.~35, no.~1, pp. 762--772, 2020.

\bibitem{hoeltgebaum2018generating}
H.~Hoeltgebaum, C.~Fernandes, and A.~Street, ``Generating joint scenarios for renewable generation: The case for non-gaussian models with time-varying parameters,'' \emph{IEEE Transactions on power systems}, vol.~33, no.~6, pp. 7011--7019, 2018.

\bibitem{5298967GMM}
R.~Singh, B.~C. Pal, and R.~A. Jabr, ``Statistical representation of distribution system loads using gaussian mixture model,'' \emph{IEEE Transactions on Power Systems}, vol.~25, no.~1, pp. 29--37, 2010.

\bibitem{NonGaussianvanin2023exact}
M.~Vanin, T.~Van~Acker, R.~D’hulst, and D.~Van~Hertem, ``Exact modeling of non-gaussian measurement uncertainty in distribution system state estimation,'' \emph{IEEE Trans. Instrumentation and Measurement}, 2023.

\bibitem{NonGaussian8008784}
P.~A. Pegoraro, A.~Angioni, M.~Pau, A.~Monti, C.~Muscas, F.~Ponci, and S.~Sulis, ``Bayesian approach for distribution system state estimation with non-gaussian uncertainty models,'' \emph{IEEE Transactions on Instrumentation and Measurement}, vol.~66, no.~11, pp. 2957--2966, 2017.

\bibitem{zayyani2024robust}
H.~Zayyani, M.~Korki, and A.~Taghavi, ``A robust stable laplace continuous mixed norm adaptive filter algorithm,'' \emph{IEEE Sensors Letters}, vol.~8, no.~3, pp. 1--4, 2024.

\bibitem{zamzam2019data}
A.~S. Zamzam, X.~Fu, and N.~D. Sidiropoulos, ``Data-driven learning-based optimization for distribution system state estimation,'' \emph{IEEE Transactions on Power Systems}, vol.~34, no.~6, pp. 4796--4805, 2019.

\bibitem{liu2025noise}
Y.~Liu and J.~Ma, ``Noise-trained deep learning-based distribution system state estimation considering the penetration of distributed energy resources,'' \emph{IEEE Transactions on Smart Grid}, 2025.

\bibitem{JZhTPWRS10184461}
J.~Zhang, J.~Zhao, J.~Yang, and J.~Zhao, ``Deep multi-fidelity bayesian data fusion for probabilistic distribution system voltage estimation with high penetration of pvs,'' \emph{IEEE Transactions on Power Systems}, vol.~39, no.~2, pp. 3661--3672, 2024.

\bibitem{azimian2022state}
B.~Azimian, R.~S. Biswas, S.~Moshtagh, A.~Pal, L.~Tong, and G.~Dasarathy, ``State and topology estimation for unobservable distribution systems using deep neural networks,'' \emph{IEEE transactions on instrumentation and measurement}, vol.~71, pp. 1--14, 2022.

\bibitem{tran2021}
M.-Q. Tran, A.~S. Zamzam, and P.~H. Nguyen, ``Enhancement of distribution system state estimation using pruned physics-aware neural networks,'' in \emph{2021 IEEE Madrid PowerTech}, 2021, pp. 1--5.

\bibitem{zamzam2020}
A.~S. Zamzam and N.~D. Sidiropoulos, ``Physics-aware neural networks for distribution system state estimation,'' \emph{IEEE Transactions on Power Systems}, vol.~35, no.~6, pp. 4347--4356, 2020.

\bibitem{yuan2022joint}
Y.~Yuan, K.~Dehghanpour, Z.~Wang, and F.~Bu, ``A joint distribution system state estimation framework via deep actor-critic learning method,'' \emph{IEEE Transactions on Power Systems}, vol.~38, no.~1, pp. 796--806, 2022.

\bibitem{caodi2024tpwps}
D.~Cao, J.~Zhao, W.~Hu, N.~Yu, J.~Hu, and Z.~Chen, ``Physics-informed graphical learning and bayesian averaging for robust distribution state estimation,'' \emph{IEEE Trans. Power Syst.}, vol.~39, no.~2, pp. 2879--2892, 2024.

\bibitem{manitsas2012distribution}
E.~Manitsas, R.~Singh, B.~C. Pal, and G.~Strbac, ``Distribution system state estimation using an artificial neural network approach for pseudo measurement modeling,'' \emph{IEEE Transactions on power systems}, vol.~27, no.~4, pp. 1888--1896, 2012.

\bibitem{huang2019probabilistic}
Y.~Huang, Q.~Xu, C.~Hu, Y.~Sun, and G.~Lin, ``Probabilistic state estimation approach for ac/mtdc distribution system using deep belief network with non-gaussian uncertainties,'' \emph{IEEE Sensors Journal}, vol.~19, no.~20, pp. 9422--9430, 2019.

\bibitem{dobbe2019linear}
R.~Dobbe, W.~van Westering, S.~Liu, D.~Arnold, D.~Callaway, and C.~Tomlin, ``Linear single-and three-phase voltage forecasting and bayesian state estimation with limited sensing,'' \emph{IEEE Transactions on Power Systems}, vol.~35, no.~3, pp. 1674--1683, 2019.

\bibitem{pse8616788}
K.~Dehghanpour, Y.~Yuan, Z.~Wang, and F.~Bu, ``A game-theoretic data-driven approach for pseudo-measurement generation in distribution system state estimation,'' \emph{IEEE Transactions on Smart Grid}, vol.~10, no.~6, pp. 5942--5951, 2019.

\bibitem{mestav2019bayesian}
K.~R. Mestav, J.~Luengo-Rozas, and L.~Tong, ``Bayesian state estimation for unobservable distribution systems via deep learning,'' \emph{IEEE Transactions on Power Systems}, vol.~34, no.~6, pp. 4910--4920, 2019.

\bibitem{raghuvamsi2023review}
Y.~Raghuvamsi and K.~Teeparthi, ``A review on distribution system state estimation uncertainty issues using deep learning approaches,'' \emph{Renewable and Sustainable Energy Reviews}, vol. 187, p. 113752, 2023.

\bibitem{abur2004power}
A.~Abur and A.~G. Exposito, \emph{Power system state estimation: theory and implementation}.\hskip 1em plus 0.5em minus 0.4em\relax CRC press, 2004.

\bibitem{blundell2015weight}
C.~Blundell, J.~Cornebise, K.~Kavukcuoglu, and D.~Wierstra, ``Weight uncertainty in neural network,'' in \emph{International conference on machine learning}.\hskip 1em plus 0.5em minus 0.4em\relax PMLR, 2015, pp. 1613--1622.

\bibitem{heckerman2022tutoriallearningbayesiannetworks}
\BIBentryALTinterwordspacing
D.~Heckerman, ``A tutorial on learning with bayesian networks,'' 2022. [Online]. Available: \url{https://arxiv.org/abs/2002.00269}
\BIBentrySTDinterwordspacing

\bibitem{Feeder}
\BIBentryALTinterwordspacing
(2017) {IEEE} test feeder specifications. [Online]. Available: \url{http://sites.ieee.org/pes-testfeeders/resources}
\BIBentrySTDinterwordspacing

\bibitem{OEDI_Dataset_4568}
\BIBentryALTinterwordspacing
C.~Deline, K.~Perry, M.~Deceglie, M.~Muller, W.~Sekulic, and D.~Jordan, ``Photovoltaic data acquisition (pvdaq) public datasets,'' Open Energy Data Initiative (OEDI), NREL, https://doi.org/10.25984/1846021, 2021. [Online]. Available: \url{https://data.openei.org/submissions/4568}
\BIBentrySTDinterwordspacing

\bibitem{winddata}
\BIBentryALTinterwordspacing
Wind Data Hub. [Online]. Available: \url{https://wdh.energy.gov/project/wind}
\BIBentrySTDinterwordspacing

\bibitem{fu2027polar}
C.~Fu, Z.~Tang, F.~Chen, Y.~Cheng, and Z.~Zheng, ``Polar graph mixture density network for probabilistic power system state estimation,'' \emph{Electric Power Systems Research}, vol. 264, p. 113848, 2027.

\bibitem{liu2019}
Y.~Liu, J.~Li, and L.~Wu, ``State estimation of three-phase four-conductor distribution systems with real-time data from selective smart meters,'' \emph{IEEE Trans. Power Systems}, vol.~34, no.~4, pp. 2632--2643, 2019.

\bibitem{LinearPF7031975}
S.~Bolognani and S.~Zampieri, ``On the existence and linear approximation of the power flow solution in power distribution networks,'' \emph{IEEE Transactions on Power Systems}, vol.~31, no.~1, pp. 163--172, 2016.

\end{thebibliography}
\let\mybibitem\bibitem

\end{document}